%% file: ms.tex
\documentclass[preprint]{aastex62}
\usepackage{natbib}
\received{Sept. 3, 2018}
\accepted{Dec. 5,  2018}

\submitjournal{ApJ}

\begin{document}

\title{CHANDRA OBSERVATIONS OF THE MASSIVE STAR-FORMING REGION ONSALA 2}

\correspondingauthor{Stephen L. Skinner}
\email{stephen.skinner@colorado.edu} 

\author{Stephen L. Skinner}
\affiliation{Center for Astrophysics and 
Space Astronomy (CASA), Univ. of Colorado,
Boulder, CO, USA 80309-0389}

\author{Kimberly R. Sokal}
\affiliation{Dept. of Astronomy, Univ. of Texas, Austin, TX 78712-1205}

\author{Manuel  G\"{u}del}
\affiliation{Dept. of Astrophysics, Univ. of Vienna, 
T\"{u}rkenschanzstr. 17,  A-1180 Vienna, Austria} 

%
\newcommand{\ltsimeq}{\raisebox{-0.6ex}{$\,\stackrel{\raisebox{-.2ex}%
{$\textstyle<$}}{\sim}\,$}}
%
\newcommand{\gtsimeq}{\raisebox{-0.6ex}{$\,\stackrel{\raisebox{-.2ex}%
{$\textstyle>$}}{\sim}\,$}}
%
\begin{abstract}
\small{
Previous radio and infrared observations have revealed an obscured
region of high-mass star formation in Cygnus known as Onsala 2 (ON 2).
Within this region lies the optically-revealed young stellar
cluster Berkeley 87 which contains several OB stars and the rare
oxygen-type Wolf-Rayet star WR 142. Previous radio studies of 
ON 2 have also discovered masers and several H II regions excited 
by embedded OB stars. Radio and GAIA parallaxes have now shown
that the H II regions are more distant than Berkeley 87.
We summarize two {\em Chandra} X-ray observations of ON 2
which detected more than 300 X-ray sources. Several
optically-identified stars in Berkeley 87 were detected
including massive OB stars and WR 142, the latter being
a faint hard source whose X-ray emission likely arises in
hot thermal plasma.  Intense X-ray emission was detected near the
compact H II regions G75.77$+$0.34 and G75.84$+$0.40 consisting 
of numerous point sources and diffuse emission. Heavily-absorbed
X-ray sources and their near-IR counterparts that may be
associated with the exciting OB stars of the H II regions are
identified. Shocked winds from embedded massive stars offer
a plausible explanation of the diffuse emission. Young
stellar object candidates in the ON 2 region are identified using near-IR
colors, but surprisingly few counterparts of X-ray sources 
have near-IR excesses typical of classical T Tauri stars.
}
\end{abstract}



\section{Introduction}

Massive young stars spend  their short pre-main-sequence
(PMS) lifetimes embedded in the dense cloud material from which they
formed and are invisible to optical telescopes. Thus, previous studies  
have traditionally focused on radio and infrared observations to probe 
the gas and dust environments of massive newborn stars. X-ray observations 
have emerged as another technique for studying young heavily-obscured 
stars since X-rays at energies of a few keV can penetrate high 
visual extinction of up to A$_{\rm V}$ $\sim$ 50 mag. X-rays
provide information on physical processes such as 
magnetic activity or shocked winds and outflows
which can produce hot plasma (T $\gtsimeq$ 10$^{6}$ K).
In addition, absorption of lower energy X-rays  below 
$\approx$1 keV by intervening gas along the line-of-sight
yields extinction estimates independent of optical measurements.

In this study we present results of sensitive {\em Chandra} 
X-ray observations of the  star-forming region Onsala 2 (ON 2), 
located in an obscured region of Cygnus. 
Several lines of evidence
point to recent and ongoing high-mass star formation in ON 2 
including radio-detected H II regions excited by young OB stars, 
a type I OH maser (Elld\'{e}r  et al. 1969; Winnberg 1970; Hardebeck \& Wilson 1971),
and a water maser (Forster et al. 1978).
Little is so far known about the population of young low-mass                                                                  \
($\ltsimeq$2 M$_{\odot}$) pre-main sequence stars, i.e. T Tauri stars,
that presumably coexists with the nascent high-mass stars.
Figure 1 shows a POSS2 red image of ON 2 with key objects identified.

Our main objectives were to (i) search for X-ray point sources in the immediate
vicinity of radio-identified H II regions that might be associated with
heavily-embedded young stars and characterize their X-ray properties,
(ii) confirm the presence of diffuse X-ray emission near known H II 
regions that was previously reported on the basis of {\em XMM-Newton} 
observations (Oskinova et al. 2010), and (iii) confirm a previous
faint X-ray detection of WR 142 (Oskinova et al. 2009; Sokal et al. 2010).
This study extends our earlier
{\em Chandra} X-ray study of optically-revealed massive stars in
the southern part of ON 2 (Sokal et al. 2010). The {\em Chandra}
data provide lower background emission, higher spatial resolution,
and more reliable source identification compared to the previous 
{\em XMM-Newton} X-ray  observation.
However, {\em XMM-Newton} provides better sensitivity to soft X-rays
at energies below $\approx$1 - 2  keV.

An optical study of the region by Turner \& Forbes (1982, hereafter TF82) 
identified several candidate members of the young open cluster 
Berkeley 87 including the rare oxygen-type Wolf-Rayet star WR 142.
A distance of 1.23 kpc to Berkeley 87
was derived by Turner et al. (2006) and this value was adopted
in our previous {\em Chandra} study (Sokal et al. 2010).
{\em Gaia} Data Release 2 (DR2) is now providing accurate
parallaxes for stars in Berkeley 87, thus removing most of
the distance uncertainty in previous studies. The {\em Gaia} DR2
parallax for WR 142 gives a distance of 1.737 ($+$0.091, $-$0.081) kpc.
For the X-ray detected Berkeley 87 OB stars listed in 
Table 2 of Sokal et al. (2010), {\em Gaia} DR2 parallaxes
give distances in the range 1.488 - 1.812 kpc, with a 
mean value near 1.75 kpc. We thus adopt 1.75 kpc as a representative
distance for Berkeley 87 in this study. An accurate distance and
dispersion based on all cluster members will be possible 
when the final {\em Gaia} data release becomes available.
Age estimates for the  Berkeley 87 cluster are in the range
$\sim$3 - 5 Myr (Massey et al. 2001; Turner et al. 2010).
We emphasize that the above age estimates and distance of $\sim$1.75 kpc 
apply only to the Berkeley 87 star cluster. The H II regions lying along 
or near the line-of-sight toward Berkeley 87 are more distant and
their embedded ionizing stars are undoubtedly much younger (Dent et al. 1988;
Tremblin et al. 2014).

The OH maser lies in  close proximity to the
ultracompact H II (ucH II) region G75.78$+$0.34, which has been the
subject of many radio continuum studies (e.g. Matthews et al. 1973,
hereafter M73; Matthews et al.  1977; Matthews \& Spoelstra 1983; Harris 1976;
Wood \& Churchwell 1989). Because of its ultracompact morphology, G75.78$+$0.34
is thought to be a massive star at a very early evolutionary  stage.
Radio trigonometric parallax measurements give a distance of
3.56 kpc to maser G75.78$+$0.34 (Xu et al. 2013), so this ucH II
region  lies twice as far away as  Berkeley 87 
and is not physically associated with the optically-revealed cluster.

Several other H II regions in ON 2 have been identified and their
positions are marked in Figure 1.
Two of these, G75.77$+$0.34 and G75.84$+$0.40, have dense groupings of X-ray 
sources in their immediate vicinity suggestive of recent star formation, as
shown below. The compact H II (cH II) region G75.77$+$0.34 
has a double-peaked radio morphology 
with a separation of $\approx$19$''$ between the radio peaks
(M73, Matthews et al. 1977; Harris 1976). 
This region emits brightly in the IR at 10.7 $\mu$m and 19 $\mu$m (Hefele et al. 1977).
Lying in the northern part of ON 2 is the cH II region G75.84$+$0.40 which 
also has double-peaked radio morphology with a separation between peaks
of $\approx$20$''$ (M73). Previous radio studies of G75.77$+$0.34 and G75.84$+$0.40
have usually assumed a distance $\sim$5 - 6 kpc based on the kinematic
distance d = 5.5 kpc to ON 2 estimated by Reifenstein et al. (1970),
but the study of Garay et al. (1993) adopted 4.1 kpc.
However, the radio parallax measurements of Xu et al. (2013) give d = 3.51 kpc for
the maser G75.76$+$0.33, suggesting that the kinematic distance of
5.5 kpc may be an overestimate. In this study, we adopt a distance of
3.5 kpc to the H II regions G75.77$+$0.34 and G75.84$+$0.40.

Radio molecular line studies have probed the rich gas environment in ON 2.
Complex $^{12}$CO line wings were seen toward
G75.77$+$0.34 by Matthews et al. (1986) and interpreted as
possible interaction between the cH II region and its surroundings.
Dense gas was mapped in NH$_{3}$ and HCO$^{\rm +}$  by Dent et al. (1988)
who found evidence for local gas heating  near the embedded
stars that excite the H II regions.
Shepherd, Churchwell, \& Wilner (1997) obtained  high spatial resolution
molecular line and millimeter continuum maps of the ON 2 core region.
They detected at least four massive molecular outflows
and a 3 mm continuum peak at G75.78$+$0.34.
Two fainter continuum sources probably associated with dust-enshrouded 
young stellar objects (YSOs) were found. 
They also noted an enhanced SiO abundance which they
hypothesized could be due to grain destruction caused by
shocks from winds or outflows impacting surrounding material.
Submillimeter mapping at 450 $\mu$m and 800 $\mu$m 
by Hunter (1998) detected several compact sources (candidate protostars)
near the H II regions.

The  {\em Chandra} observations discussed here reveal dense groupings of
X-ray point sources near G75.77$+$0.34 and G75.84$+$0.40, some of which 
are likely to be young stars.
We  find evidence for diffuse X-ray
emission near these H II regions (see also Oskinova et al. 2010) 
with spectral properties suggestive of a thermal origin. Diffuse
X-ray emission interpreted as thermal emission from shocked stellar
winds has also been detected by {\em Chandra} in other massive star-forming 
regions including the Arches cluster (Yusef-Zadeh et al. 2002), and
M17 and the Rosette Nebula (Townsley et al. 2003).  Diffuse
thermal X-ray plasma in the Orion Nebula region excited by
massive Trapezium stars was also reported on the basis of
{\em XMM-Newton} observations by G\"{u}del et al. (2008). In contrast, 
the diffuse X-ray emission detected by {\em Chandra} in RCW 38 has been interpreted
as nonthermal (Wolk et al. 2002) and diffuse emission in Westerlund 1
is consistent with either a nonthermal or hot thermal plasma 
origin (Muno et al. 2006). We  confirm here the previous detection 
of WR 142 as a hard X-ray source and characterize the emission of the 
brightest X-ray source in ON 2, which has no cataloged optical counterpart.

\section{Observations and Data Reduction}

A {\em Chandra} observation of the southern portion of Onsala 2
(ON2S) was obtained in 2009 with pointing centered near WR 142
which lies about 3$'$ south of G75.77$+$0.34 (Fig. 2; Sokal et al. 2010). 
The northern portion (ON2N) was observed in 2016 with pointing
centered $\approx$1$'$ south of G75.84$+$0.40.
Both observations used the ACIS-I (Advanced CCD Imaging Spectrometer) 
imaging array which captures a $\approx$16.9$'$ $\times$ 16.9$'$
field-of-view (FoV). There was some overlap in the FoV between
the two exposures as shown in Figure 2.  Table 1 summarizes
the observations and ACIS-I detector properties.

Using the  Portable Interactive Multi-Mission
Simulator (PIMMS), the intrinsic (unabsorbed) X-ray luminosity 
detection limits for an on-axis point source and a 5-count 
detection threshold (0.3 - 8 keV) were 
log L$_{x}$ = 29.79 ergs s$^{-1}$ (ON 2S) 
and log L$_{x}$ = 29.88 ergs s$^{-1}$ (ON 2N) at a characteristic distance of
1.75 kpc. These values assume a generic T Tauri star thermal
plasma spectrum with  kT $\approx$ 2-3 keV and absorption column 
density N$_{\rm H}$ $\approx$ 10$^{22}$ cm$^{-2}$. This absorption 
corresponds to A$_{\rm V}$ $\approx$ 4.8 mag, a typical value
for optically-revealed Berkeley 87 stars (TF82; Sokal et al. 2010).
For accreting classical T Tauri stars (cTTS) the above  L$_{x}$
limits correspond to stellar mass detection limits in the range
M$_{*}$ $\approx$ 0.4 - 0.6 M$_{\odot}$
based on established correlations (Preibisch et al. 2005;
Telleschi et al. 2007). 
For the more distant H II regions at $\sim$3.5 kpc, the
detection limits are higher and correspond to more
massive cTTS with M$_{*}$ $>$ 1 M$_{\odot}$ (Secs. 4.1 and 4.3).

Data analysis was based on  standard processing files  provided 
by the {\em Chandra}  X-ray Center (CXC) and followed standard
CXC processing threads using version 4.9 of the 
{\em Chandra} Interactive Analysis of Observations (CIAO)
software. X-ray source positions, 3$\sigma$ position error ellipses,
and net counts (background-subtracted)  were obtained using the 
CIAO {\em wavdetect} tool with observation-specific point-spread-function
(PSF) images applied to energy-filtered event files.
{\em Wavdetect} was run separately for each of the four ACIS-I
CCDs using the input value {\em falsesrc} = 1.0 to assure no
more than one false source per scale size. Each run was executed
using four scale sizes (wavelet radii): 1,2,4, and 8 pixels with
a pixel size of 0$''$.492.
We thus anticipate no more than four false sources per ACIS-I CCD
in each observation.
Checks for X-ray variability were made using the
CIAO {\em glvary} tool (Gregory \& Loredo 1992, 1996).
For brighter sources, spectra and associated instrument
response files were extracted using CIAO {\em specextract}.
Spectra were fitted using HEASARC's {\em XSPEC} fitting package 
(v. 12.8.2) to obtain estimates of absorption column
density (N$_{\rm H}$), mean plasma energy (kT) for thermal models, 
photon power-law index ($\Gamma_{\rm ph}$) for power-law models, 
normalization (norm), and
X-ray flux F$_{x}$ (absorbed and unabsorbed). 
We used the 
CIAO {\em Image of Diffuse Emission} thread 
to search for evidence of diffuse emission near the cH II regions (Sec. 3.8).

\input{table1.tex}

\section{Results}      
   
\subsection{X-ray  Sources in ON 2 }

Figure 2-left is a mosaic of the two {\em Chandra} images with
X-ray source positions overlaid. 
In total, 201 sources were detected
in ON 2N and 209 sources in ON 2S. A comparison of the 
X-ray positions of the sources in the two fields reveals 
34 sources in the overlap region whose
R.A. and declination  differ by $\leq$1$''$. These
sources are probable duplicate detections.
An additional 8 sources in the overlap region
have R.A. and declination values which differ by
1$''$ - 1$''$.5 and are possible matches.

The sources are widely distributed over
the two fields but the most intense X-ray emission is
concentrated near the cH II region G75.77$+$0.34,
as apparent from the smoothed ACIS image in 
Figure 2-right and the source list in Table 2. 
A second less conspicuous grouping of 
sources is present to the north near G75.84$+$0.40 (Table 3).
The brightest point source is CXO J202152.99$+$372139.33 in ON 2S.
No optical counterpart was found for this source but
it is coincident with a faint IR source (Sec. 3.7).
The brightest X-ray point source with an identified optical 
counterpart is the B1 Ia star HD 229059 (Sokal et al. 2010),
a Berkeley 87 member. 
The oxygen-type WR star WR 142 was detected in both 
{\em Chandra} observations. We give additional details below 
on the X-ray emission near the H II regions and other
sources of interest in ON 2.

\subsection{H II Region G75.77+0.34}

Figure 3 shows the {\em Chandra} image in a zoomed 
1$'$ $\times$ 1$'$ region centered on G75.77$+$0.34
with positions of all cataloged 2MASS sources overlaid.
Fifteen X-ray sources with at least 5 net counts were detected 
in this region (Table 2).  
All fifteen X-ray sources have a UKIDSS Galactic Plane Survey 
(UGPS; Lucas et al. 2008) near-IR counterpart and
nine have 2MASS counterparts (Table 2). 
Because of the tight X-ray source clustering, we suspect
that some of these sources are YSOs. But, as we discuss further
in Section 4.1, evidence for near-IR excesses indicative of
disked cTTS is generally lacking.
Table 2 undoubtedly contains objects that are not young
stars such as field stars and extragalactic background sources.
Any field stars would likely be foreground stars since 
background star contamination is expected to be low because
of the  high extinction toward this region (see below).
Some of the sources in Table 2 are classified as galaxies
in the UGPS catalog. These source classifications are
based on the source image profile in the UKIRT Wide Field Camera 
Science Archive  (Hambly et al. 2008) and not on spectra.
Marginally-resolved pairs of 
stars in crowded fields such as we are studying here
can be erroneously classified as galaxies (Lucas et al. 2008).
Thus, we make very limited use of the UGPS source classes in
this study.

The Westerbork radio map of G75.77$+$0.34 at 5 GHz shows two radio peaks
separated by 18.$''$9 identified as
G75.77$+$0.34A and G75.77$+$0.34B (M73).
Radio peak A is offset by less than 1$''$ from a
K$_{s}$ = 12.5 mag 2MASS source (2M J202142.13$+$372554.3)
which is also faintly visible in {\em Spitzer} IRAC
3.6 $\mu$m images (Fig. 4). However, no {\em Chandra}
source is detected at the position of radio peak 
G75.77$+$0.34A. 
The Very Large Array (VLA) radio position of G75.77$+$0.34
given by Garay et al. (1993) is offset slightly to the west of
the G75.77$+$0.34A Westerbork position. There is no {\em Chandra} 
source at the VLA position; the nearest {\em Chandra} source 
(CXO J202141.83$+$372550.0) is displaced 2$''$.5 to the south.
In contrast, the radio position error box of G75.77$+$0.34B
overlaps the {\em Chandra} source CXO J202140.54$+$372548.3 (Fig. 3; Table 2) 
as well as a 2MASS near-IR source (2MASS J202140.61$+$372548.4) and 
a UKIDSS GPS near-IR source (UGPS J202140.54$+$372548.4).
The 2MASS source is only detected in one band (K$_{s}$ = 12.65 $\pm$ 0.07)
and the confusion flag is set. The UGPS source is detected at
J, H, and K with K = 13.366 $\pm$ 0.003 and is classified as a
star in the UGPS catalog. The 2MASS and UGPS sources are probably
the same object but the K$_{s}$ and K magnitudes are discrepant.
The UGPS position is offset by only 0$''$.15 from the {\em Chandra} 
position so there is little doubt that the UGPS source is the near-IR
counterpart of the {\em Chandra} source.
The {\em Spitzer} IRAC 3.6 $\mu$m image in Figure 4
shows nebulosity at radio peak B but no discernible point source 
within the  position error box of G75.77$+$0.34B.
This nebulosity fills the region between radio peaks A and B
which includes several near-IR sources (Fig. 3).

The  Westerbork radio position of G75.77$+$0.34B from M73
converted to equinox J2000 is
J202140.54$+$372550.2, which agrees with
the position of  X-ray souce CXO J202140.54$+$372548.3
in R.A. but is offset northward  in DEC by  1$''$.9.
We note here that the Westerbork telescope radio position
of the ultracompaxt H II region G75.78$+$0.34 (M73) is offset
0.$''$7 northward of the  {\em VLA} position determined
by Wood \& Churchwell (1989). The above comparisons suggest
that a slight systematic northward shift of the Westerbork 
telescope radio positions by $\approx$0$''$.7 may be
present. If that is the case then the offset 
between the Westerbork radio position of
G75.77$+$0.34B and  {\em Chandra} source
position is reduced to 1$''$.2, making the
correspondence between the two very likely.

The {\em Chandra} spectrum of the source lying inside the 
G75.77$+$0.34B radio position error box  
is shown in Figure 3-bottom. The spectrum is heavily absorbed below 2 keV.
A fit of the spectrum with a simple  1T $apec$ thermal plasma model 
is summarized in Table 4 and overlaid in Figure 3.
The 1T $apec$ fit gives an absorption 
N$_{\rm H}$ = 5.06 [3.65 - 7.07; 1$\sigma$] $\times$ 10$^{22}$ cm$^{-2}$.
This value is $\approx$5 times larger than obtained in similar
fits of bright OB stars detected by {\em Chandra} in 
Berkeley 87 (Sokal et al. 2010).
Converting the above X-ray absorption  to 
A$_{\rm V}$ using the average 
N$_{\rm H}$ = 1.9 $\times$ 10$^{21}$ cm$^{-2}$A$_{\rm V}$ 
of the slightly different conversion factors of Gorenstein (1975)
and Vuong et al. (2003)
gives A$_{V}$ = 26.6 [19.2 - 37.2; 1$\sigma$] mag.
This corresponds to A$_{K}$ = 2.9 [2.1 - 4.1] mag
using  R$_{\rm V}$ = 3.1 and A$_{\rm K}$ = 0.34E(B$-$V)  (Bessell \& Brett 1988).

The source UGPS J202140.54$+$372548.4  associated with the {\em Chandra} 
source has observed magnitudes 
J = 15.737$\pm$0.005, H = 14.148$\pm$0.003, and K = 13.366$\pm$0.003.
If it is dereddened using the above value  A$_{K}$ = 2.9 [2.1 - 4.1] mag
then its absolute magnitude is M$_{\rm K}$ = $-$2.25 [$-$3.45 - $-$1.45]
at d = 3.5 kpc. This corresponds to a $\approx$B1V [O8.5V - B2V] star
(Martins \& Plez 2006; Mamajek 2018). Dereddening the 2MASS 
value K$_{s}$ = 12.65 gives $\approx$B0V [O6V - B1V].
Using radio continuum data, M73 derived a spectral type
of O9.5V at their adopted distance of 5.5 kpc.
Scaling the excitation parameter for G75.77+0.34B given in M73 
down to  d = 3.5 kpc gives a 
later radio-derived type of B0V or a O9.5 ZAMS star (Churchwell \& Walmsley 1973).
Thus, the spectral type determined from dereddening the K and K$_{s}$ magnitudes
of the near-IR sources using the extinction estimated from {\em Chandra} spectral 
absorption is consistent with radio determinations, but the inferred type 
is  distance-dependent.

 Although the dereddened K$_{s}$ and K magnitudes and radio data 
     point to a late O or early B-type star at the assumed distance of 3.5 kpc,
     the dereddened colors determined from
     UGPS photometry and normal extinction laws (Rieke \& Lebofsky 1985),
     in combination with the X-ray derived A$_{\rm V}$, are not consistent
     with a {\em single} late O or early B-type main sequence star. 
     The intrinsic (dereddened) colors of late OV or early BV stars 
     have small negative values of  (J$-$H)$_{0}$ = $-$0.16 - $-$0.11
     and (H$-$K)$_{0}$ = $-$0.10 - $-$0.03 (Mamajek 2018; Martins \& Plez (2006),
     but the dereddened colors of UGPS JJ202140.54$+$372548.4 have larger
     negative values if the extinction A$_{\rm V}$ = 26.6[19.2 -37.2] mag
     determined from {\em Chandra}  spectral absorption is adopted.
     In order to bring the dereddened colors and
     absolute magnitude M$_{K}$ of this UGPS source into agreement with
     a {\em single} late OV or early BV star, its extinction would need to 
     be A$_{\rm V}$ $\approx$ 12.5 - 16 mag and d $\gtsimeq$ 4 kpc. 
     Such a lower extinction could be reconciled
     with the {\em Chandra} value if excess X-ray spectral absorption is present
     above that expected for normal reddening. The most likely explanation for such
     absorption would be cold gas along the line-of-sight. Such X-ray absorbing
     gas does not contribute significantly to optical extinction and is
     often  seen in protostars and in massive stars with dense winds such as  
     WR stars (e.g. Skinner et al. 2010). Other factors that could
     lead to anomalous dereddened colors of the UGPS source include
     unresolved binarity. 

\input{table2.tex}

\clearpage

\subsection{H II Region G75.84$+$0.40}
The {\em Chandra} image near G75.84$+$0.40
is shown in Figure 5. Fourteen X-ray sources
with at least 5 net counts
were detected in a 1$'$ $\times$ 1$'$ region
centered on G75.84$+$0.40 (Table 3), but only
four have a 2MASS or UKIDSS GPS near-IR counterpart.
The low fraction of X-ray sources with identifiable
near-IR counterparts is mainly due to the high near-IR
background emission in this region coming from
the IR-bright (K$_{s}$  = 0.3) M3 supergiant BC Cyg 
located 48$''$ to the north (Fig. 1). The bright emission 
from BC Cyg  also adversely affects {\em Spitzer} IRAC 
images in this region. A search of the {\em Spitzer}
Enhanced Imaging Products (SEIP) archives reveals
a mid-IR counterpart for only one of the X-ray sources 
in Table 3, CXO J202138.01$+$373114.5. It is also
the only source in Table 3 with a 2MASS match.
Near-IR colors needed to identify IR excess sources
are lacking. The most likely YSO candidates are
the bright hard  variable source
CXO J202138.30$+$373124.2 and the two sources
near G75.84$+$0.40B discussed below.

There is no {\em Chandra} source
coincident with Westerbork radio peak G75.84$+$0.40A (M73)
or with the VLA position of G75.84$+$0.40 given by Garay et al. (1993).
However, the radio position error box of  Westerbork peak G75.84$+$0.40B 
overlaps the position error ellipse of
CXO J202137.69$+$373115.1, even if the possible
0.$''$7 northward systematic offset in the Westerbork
radio position is taken into account.  
There is no 2MASS or UGPS source
within the {\em Chandra} error ellipse or the 
position error box of radio peak B but a faint {\em Spitzer}
IRAC source is visible (IRAC1 J202137.68$+$373115.6)
and nearly centered within the {\em Chandra} error ellipse
(Fig. 4-bottom).
In addition, a brighter {\em Chandra} source CXO J202138.01$+$373114.5
with a K$_{s}$ = 9.628$\pm$0.029  2MASS counterpart 
(2MASS J202137.97$+$373115.2) lies just 2$''$.6 
east of radio peak B. 
This IR source is not found in the UGPS catalog
but is clearly visible in 2MASS and IRAC 
images (Fig. 4-bottom; IRAC1 J202137.97$+$373115.5).

X-ray spectra  were extracted for both of the above
sources, i.e. CXO J202137.69$+$373115.1 and 
CXO J202138.01$+$373114.5.
The spectra are shown in Figure 5 and spectral
fits  are summarized in Table 4.
The X-ray source lying closest to radio peak B
is fainter and slightly harder, i.e. a larger fraction of its
X-ray events emerge in the hard 2-8 keV band.
The second brighter X-ray source lying 2$''$.6 east
shows a feature near 1.85 keV that may include emission
from the Si XIII triplet which has maximum emissivity
at $\approx$10 MK. Neither of these
two sources is as hard or as luminous in X-rays as the 
{\em Chandra} source identified with the
cH II region  radio peak G75.77$+$0.34B.

X-ray spectral fits of the above two sources with isothermal models
require high temperatures and similar absorption values
equivalent to A$_{\rm V}$ $\approx$ 9 - 10 mag, but uncertainties
in the derived spectral parameters are large due to the limited
number of counts (Table 4). For the brighter X-ray source
CXO J202138.01$+$373114.5 with a 2MASS counterpart, the 
simple isothermal model gives an absorption 
N$_{\rm H}$ = 1.63 [1.01 - 2.71; 1$\sigma$] $\times$ 10$^{22}$ cm$^{-2}$
which equates to A$_{\rm V}$ = 8.6 [5.3 - 14.3] mag.
The inferred K-band extinction is then A$_{\rm K}$ = 0.94 [0.58 - 1.57].

The near-IR  source 2MASS J202137.97$+$373115.2 can be dereddened 
using the above  A$_{\rm K}$ value.  But it should be noted 
that the annular region used to determine background in the
2MASS automated processing captures part of a bright diffraction
spike from BC Cyg to the north and part of the emission from
a second source lying $\approx$15$''$ to the southeast.  
As such, the background is modestly overestimated and the 2MASS source is 
slightly brighter than its assigned K$_{s}$ = 9.628$\pm$0.029 magnitude.
Also, there is no cataloged UGPS counterpart to use as a cross-check 
on the 2MASS K$_{s}$ magnitude. 
Keeping the above in mind and proceeding as above (Sec. 3.2), 
the 2MASS value K$_{s}$ = 9.628 mag converts to 
K$_{s,dered}$ = 8.68 [8.06 - 9.05] and 
M$_{\rm K}$ = $-$4.04 [$-$4.66 - $-$3.67] at d = 3.5 kpc.
This corresponds to a O6.5V [O4V - O7.5V] star (Martins \& Plez 2006).

For comparison, the radio-derived spectral type of 
the source G75.84$+$0.40B is O8V (M73; d = 5.5 kpc).
Scaling this down to our adopted distance of 3.5 kpc 
gives a O9.5V or ZAMS O9 star (Churchwell \& Walmsley 1973).
The above difference in spectral types determined from the
2MASS and Westerbork radio sources along  with  their positional
offset and the presence of multiple IR sources (Fig. 4-bottom)
raises the possibility that more than one heavily-reddened 
high-mass star is present near radio peak G75.84$+$0.40B.

\input{table3.tex}

\clearpage

\input{table4.tex}

\subsection{Ultracompact H II Region G75.78$+$0.34}

Figure 6-left shows the {\em Chandra} image near G75.78$+$0.34.
No X-ray source was detected by {\em wavdetect} at the  
VLA radio position given by Shepherd et al. (1997).
But a faint X-ray peak is visible in the merged 
image at an offset 2$''$.8 southwest of the radio position.
This source (CXO J202143.92$+$372636.6) has only 9 net counts
and is very hard with H.R. = cts(2-8 keV)/cts(0.2-8 keV) = 0.91,
so likely viewed through high absorption. A faint 
K = 18.48 UKIDSS GPS source (UGPS J202143.87$+$372637.7)
is cataloged at an offset of 1$''$.26 from the X-ray position.
This UGPS source was not detected at J or H bands.

In addition to G75.78$+$0.34, three
other radio continuum sources are listed in Table 2 of 
Shepherd et al. (1997). No significant X-ray emission
is seen at their radio positions but a faint
6 count {\em Chandra} source CXO J202144.73$+$372642.6
lies 2$''$.1 northeast of the radio source
VLA J202144.58$+$372641.6. The lack of X-ray
detections of these radio sources suggests 
that the radio sources are heavily-obscured.

\subsection{H II Region G75.84$+$0.36}

This H II region was detected in 610 MHz radio
observations by Matthews \& Spoelstra (1983)
who derived a ZAMS spectral type of O8 for
the ionizing source, assuming d = 2.6 kpc.
The radio position given by Matthews \& Spoelstra
converted to equinox 2000 is
J2021498$+$373011 with an uncertainty
of $\pm$5$''$ in R.A. and $\pm$8$''$ in declination.
A JCMT 850 $\mu$m continuum source is
cataloged at J202150.6$+$373010 by
Di Francesco et al. (2008). 

The closest X-ray source to the above
radio positions is CXO J202150.30$+$373014.09,
lying 6$''$.5 northeast of the Matthews \& Spoelstra
radio position (Fig. 6-right). This X-ray
source is variable and associated with
UGPS J202150.25$+$373013.7 (K = 13.803$\pm$0.004) 
and its match 2MASS J202150.25$+$373014.1 (K$_{s}$ = 13.837$\pm$0.094).
However, neither of the radio  positions
given above falls inside the position error
ellipse of this {\em Chandra} source so
no definitive association of the X-ray 
source with the radio sources can be claimed.

\subsection{WR 142}

WR 142 was detected in both {\em Chandra} observations with
nearly identical exposure times. 
But as Figure 1 shows, it was placed near the center of the ACIS-I 
CCD array in the 2009  observation (ObsId 9914) and far 
off-axis near the edge of the array in 2016 (ObsId 18083).
Since ACIS-I effective area decreases with increasing
off-axis distance, fewer net counts were obtained in
2016 (26 $\pm$ 7 counts) than in 2009 (46 $\pm$ 7 counts).

The spectrum in both observations is quite hard with
almost all detected events having energies above 1.5 keV (Fig. 7).
Low-energy absorption is anticipated based on the high visual extinction toward
Berkeley 87 cluster members (TF82; Barlow \& Hummer 1982).
The visual extinction for WR 142 given in van der Hucht (2001)
is A$_{\rm V}$ = 6.07 (A$_{v}$ = 6.74) mag, which equates to
N$_{\rm H}$ = (0.97 - 1.33) $\times$ 10$^{22}$ cm$^{-2}$
using standard conversions (Gorenstein 1975; Vuong et al 2003). 
The 2009 ACIS-I spectrum did not show any recognizable line
features. However, peaks are
visible in the 2016 spectrum at energies near well-known
high-temperature emission lines including the  Si XIII 
triplet (1.855 - 1.865 keV; maximum emissivity temperature log T$_{max,{\rm SiXIII}}$ = 7.0 K),
Ca XX (4.107 keV; log T$_{max,{\rm CaXX}}$ = 7.7 K), and the 
Fe K complex dominated by Fe XXV lines (6.67 - 6.68 keV; log T$_{max,{\rm FeXXV}}$ = 7.8 K).
These features, albeit faint, strongly suggest  that the X-ray emission of WR 142 
is thermal and dominated by hot plasma. The Si XIII triplet and Fe K emission are
commonly detected in WR stars (e.g. $\gamma^2$ Vel; Skinner et al. 2001).
Calcium lines such as Ca XIX near 3.9 keV have also been observed
in WR spectra (e.g. Schild et al. 2004; Zhekov, Gagn\'{e}, \& Skinner 2011).
The above lines are not present in a background spectrum extracted from
a large $r$ = 60$''$ circular source-free region near WR 142. The
only line-like feature in the background spectrum is near 2.1 keV and
due to the fluorescent Au M line complex excited by high-energy particles
impacting gold-plated ACIS surfaces (Bartalucci et al. 2014).

We have fitted the 2009 and 2016 ACIS-I spectra simultaneously
using an absorbed isothermal thermal plasma model (1T $apec$) and for comparison
an absorbed power-law model plus a fixed-width Gaussian line 
near 6.7 keV. For the $apec$ model we compared a solar abundance
fit with that obtained by fixing the abundances at generic
WO-star values (van der Hucht et al. 1986).
As Table 5 shows, the models give nearly identical values of
the reduced $\chi^2$ fit statistic as a result of the low number of
counts in the spectra. As such, the spectra are not of sufficient
quality to distinguish between thermal and nonthermal emission
based on the fit statistic alone but spectral peaks near known
emission lines in the 2016 spectrum tip the balance in favor of
thermal emission, at that epoch. All three models 
require substantial absorption with best-fit values in
the range N$_{\rm H}$ = (5.3 - 10.8) $\times$ 10$^{22}$ cm$^{-2}$ 
(A$_{\rm V}$ $\approx$ 28 - 57 mag) and thermal models converge to
high plasma temperatures  kT $\approx$ 5 keV.  
Confidence intervals for the best-fit parameters (Table 5) are quite 
large due to low-count statistics.
The absorption determined from the {\em Chandra} spectrum
is significantly higher than expected
on the basis of the optically-determined A$_{\rm V}$. 
The excess absorption seen in X-rays is very likely due to
the powerful WR 142 wind (see also Sec. 4.2).

The observed (absorbed) X-ray flux measured in the lower signal-to-noise 2016 
ACIS-I spectrum differs by  less than 30\% from that measured in the  2009 spectrum 
(Table 3 of Sokal et al. 2010).
At the {\em Gaia} DR2  distance of 
1.737 kpc the models give log L$_{x}$(0.3-8 keV) = 31.03 - 31.32 ergs s$^{-1}$,
at the low end of the range for X-ray detected  WR stars (Fig. 5 of Sokal et al. 2010).
We note that the above L$_{x}$ range is a factor of $\approx$2
larger than obtained by Sokal et al. (2010) where we adopted d = 1.23 kpc.

For comparison with {\em Chandra},  the {\em XMM-Newton}
X-ray spectrum of WR 142 is also hard and dominated by emission
above 1 keV (Fig. 5 of Oskinova et al. 2009). In addition, the 
{\em XMM-Newton} EPIC MOS spectrum shows a soft component below 
$\approx$0.6 keV, but this component is not present in the 
EPIC $pn$ spectrum or in the {\em Chandra} ACIS-I spectra.
The absence of such soft emission in the ACIS-I spectra may
be partially due to the low ACIS-I effective area
at energies below $\approx$0.6 keV where {\em XMM-Newton}
has much better sensitivity. The observed X-ray flux of WR 142
based on {\em XMM-Newton} spectra given by Oskinova et al. (2009)
is F$_{x}$ = 4($\pm$2) $\times$ 10$^{-14}$ ergs cm$^{-2}$ s$^{-1}$.
This value is about a factor of two larger than our ACIS-I 
measurements (Table 5), but is consistent with the ACIS-I
values when the 50\% {\em XMM-Newton} flux 
uncertainty and $\approx$20\% {\em Chandra} flux uncertainty
are taken into account.  A somewhat higher measured flux would  
be expected for {\em XMM-Newton} since it includes soft emission 
that was not detected by {\em Chandra}.

Further discussion of the origin of X-rays in WR 142 and WR stars as a 
class is given in Section 4.2.  

\input{table5.tex}

\clearpage

\subsection{The Bright X-ray Source CXO J202152.99$+$372139.33}

This is the brightest {\em Chandra} source detected in ON 2
with 498 net counts (0.2-8 keV; ObsId 9914). By comparison, the 
second brightest source is the B1 supergiant HD 229059 
with 339 net counts (ObsId 9914; Sokal et al. 2010). 
No optical, radio, or 2MASS counterpart was found within the 3$\sigma$ 
X-ray position error circle (radius = 1$''$.3) using the HEASARC {\em Browse}
search tool. This search included multiple optical and radio data
bases. Furthermore, no {\em Gaia} DR2 counterpart was found.
However, the object was faintly detected in the UKIDSS GPS 
survey  at K = 17.630$\pm$0.121 mag (UGPS J202152.99$+$372139.3) 
but no J or H band detections were reported.
The UKIDSS GPS source classification is
stellar with high probability (P$_{*}$ = 0.90).

The source is also cataloged as a 
{\em Spitzer} IRAC detection (SSTSL2 J202152.98+372139.2) and is 
visible in archived IRAC images (Fig. 8) but not in {\em WISE} images.
IRAC magnitudes computed from flux densities given in the IRSA {\em Spitzer} source
catalog  are I1 = [14.99], I2 = [13.93], I3 = [13.03], I4 = [12.27], and
a 3$\sigma$  MIPS 24$\mu$m limit M1 $>$ [9.80]. The spectral energy
distribution $\lambda$F$_{\lambda}$ versus $\lambda$ peaks in the I3
band near 5.8 $\mu$m (Fig. 8). IRAC colors are consistent with a 
class I protostar or a heavily-reddened class II YSO (Gutermuth et al. 2008).

The X-ray source is very hard with a hardness ratio = 
counts(2-8 keV)/counts(0.2-8 keV) = 0.91. The X-ray color
system based on counts in three different energy bands 
adopted by Wang et al. (2016) gives colors $CMS$ = 0.08
and $CHM$ = 0.83, placing this source in the region occupied
by the hardest point sources in the {\em Chandra} ACIS 
source catalog (Fig. 8 of Wang et al. 2016).
A test for variability using CIAO {\em glvary}
gives a low variability probability
P$_{var}$ = 0.06 (0.2 - 8 keV events).
The X-ray spectrum (Fig. 8) shows that almost all detected
events have energies above 2 keV.
Attempts to fit the ACIS-I spectrum with absorbed thermal
{\em XSPEC} plasma models ($apec$, $bremss$) require very
high and physically unrealistic temperatures.
A fit using an absorbed power-law model
gives reasonably good results (Fig. 8)
and converges to N$_{\rm H}$ = 3.36 $\times$ 10$^{22}$ cm$^{-2}$
and a photon power-law index 
$\Gamma_{ph}$ = 1.15 [0.92 - 1.39, 1$\sigma$],
and observed (absorbed) flux 
F$_{x}$(0.3 - 8 keV) = 1.97 (1.70 - 2.04; 1$\sigma$)$\times$ 10$^{-13}$ ergs cm$^{-2}$ s$^{-1}$.
The above N$_{\rm H}$ equates to A$_{\rm V}$ $\approx$ 15 - 21 mag
using standard conversions (Gorenstein 1975; Vuong et al. 2003).
It is thus clear that this bright source is not a foreground object.
If it lies at the distance of the Berkeley 87 cluster ($\sim$1.75 kpc),
then the unabsorbed flux
F$_{x,unabs}$(0.3 - 8 keV) = 3.31 $\times$ 10$^{-13}$ ergs cm$^{-2}$ s$^{-1}$
equates to  log L$_{x}$(0.3 - 8 keV) = 32.08 ergs s$^{-1}$.
This value is higher than  observed for T Tauri 
stars in well-studied massive star-forming regions such as
the Orion Nebula Cluster (Preibisch et al. 2005) and is 
more typical of intermediate-mass stars (Stelzer et al. 2005).    

This bright source was also detected in the 2008 May 28-29 {\em XMM-Newton}
observation of ON 2 (ObsId 0550220101; 65 ks exposure). After
removing intervals of high background the EPIC MOS detectors
yield 541 (MOS1) and 533 (MOS2) net counts (0.2 - 8 keV) for
this source. X-ray light curves show no significant
variability and fits of the MOS spectra give similar
results as the ACIS-I fits above but a somewhat higher observed
flux  F$_{x}$ = 2.76 $\times$ 10$^{-13}$ ergs cm$^{-2}$ s$^{-1}$.
The higher flux may be at least partially due to the better sensitivity
of {\em XMM-Newton} at low energies below $\approx$1 - 2 keV where 
{\em Chandra}  detected very few counts. 
Based on X-ray properties determined from automated analysis of sources in
the {\em XMM-Newton} Serendipitous Source Catalog, the classification 
scheme of Lin et al. (2012) categorized this object as a vaguely-defined
``Galactic Plane Source'' (GPS). 
We note here that the Galactic coordinates of the source  determined from
the {\em Chandra} position are ($l$,$b$) = (75.73$^{\circ}$, $+$0.27$^{\circ}$). 

In summary, the IRAC colors are consistent with a class I protostar
or a heavily-reddened class II source. The IRAC color-color cuts derived by 
Gutermuth et al. (2008)  do not unambiguously rule out a AGN  
but the high {\em Chandra}  count rate and {\em XMM-Newton} hardness
ratios (X-ray color-color diagrams) of Lin et al. (2012) argue against
a AGN classification, 
as does the UGPS near-IR counterpart classification as a star.  
It is not yet known if this object is physically associated 
with the Berkeley 87 cluster, or lies further away.
In either case, its high L$_{x}$ 
in combination with known L$_{x}$ $\propto$ M$_{*}$ relations (Preibisch et al. 2005)
imply a rather massive object of M$_{*}$ $\gtsimeq$2 M$_{\odot}$.

\subsection{Diffuse X-ray Emission }

Images and spectra of possible diffuse emission near
G75.77$+$0.34 and G75.84$+$0.40 are discussed below.
Additional comments on possible origins of the emission
are given in Sec. 4.3.

\noindent \underline{G75.77$+$0.34} 
The ACIS-I image near G75.77$+$0.34 shows a crowded region
with several closely-spaced 
point sources (Fig. 3), as well as emission between  sources
at levels well above the background measured in source-free
regions on the same CCD distant from G75.77$+$0.34.
To further investigate the possibility of diffuse emission 
we extracted a residual image with point
sources removed using the CIAO {\em Image of Diffuse Emission}
thread. In this procedure, emission inside {\em wavdetect} point-source
3$\sigma$  position ellipses is expunged and replaced
with  background emission. The background level
is computed from the mean of a region surrounding the 
point-source ellipse.
If the position ellipse of a nearby source overlaps the background
region then the overlapping portion is excluded from the background 
calculation to avoid overestimating the background.

Figure 9-left shows a smoothed diffuse emission image of
the region near G75.77$+$0.34. The emission is concentrated
in an elongated region spanning $\approx$80$''$ along a NE-SW 
direction from G75.77$+$0.34. An arc-shaped or semi-circle
morphology is apparent.   We extracted a spectrum from an elliptical
region with semi-axes 23$''$ $\times$ 12$''$ (area = 867 arcsec$^{2}$)
lying just south of 
G75.77$+$0.34 (Fig. 9-left). This region encloses emission
well above the nominal background level. One X-ray source
(CXO J202140.75$+$372533.4; 60 net cts) was detected 
inside the extraction ellipse but its
emission was excluded during the spectral extraction.
This  diffuse emission extraction region contains a total of
241 counts (0.2 - 8 keV), of which 81$\pm$6 counts are
estimated to be due to nominal background, leaving
160$\pm$6 net counts (0.2 - 8 keV) above background.

The diffuse spectrum is shown in Figure 9-bottom
with a comparison nominal background spectrum extracted 
on the same CCD.
The diffuse emission spectrum is elevated
substantially above background in the 2 - 7 keV range
and clearly shows  line-like features near 4, 5, and 6.7 keV.
A Gaussian fit of the low-energy line gives
E = 4.01 [3.90 - 4.10; 90\% confidence] keV
which is probably Ca XX (E$_{lab}$ = 4.107 keV;
log T$_{max,{\rm CaXX}}$ = 7.7 K) but the 
feature is sufficiently broad to allow for additional
emission  from Ca XIX (E$_{lab}$ = 3.902 keV;
log T$_{max,{\rm CaXIX}}$ = 7.5 K).
The line near 5 keV has a fitted energy
E = 5.26 [5.03 - 5.54; 90\% conf.] keV
which is most likely Ca XX (E$_{lab}$ = 5.13 keV).
The best-fit energy of the Fe line is 6.63 keV so
this is Fe K emission (log T$_{max,{\rm FeXXV}}$ = 7.8 K)
and not fluorescent Fe which is sometimes seen
at a lower energy near 6.4 keV.
The presence of these  emission lines is a good indication 
that the  diffuse emission
arises in very hot thermal plasma.

A fit of the binned diffuse spectrum with a
one-temperature solar abundance thermal plasma model (1T $apec$) 
is summarized in Table 6. The fit gives a high temperature kT = 5.4 keV
and the best-fit N$_{\rm H}$
equates to A$_{\rm V}$ $\approx$ 18.9 [14.7 - 25.8] mag.
The solar abundance  1T $apec$ model begins to reproduce the Fe K
feature near 6.7 keV but does not recover the features near
4.1 and 5.2 keV.  Adding a
second temperature component does not improve
the fit. Allowing the Ca and Fe abundances to 
vary reproduces the 4.1 keV Ca XX line and the Fe K line
but not the 5.2 keV feature. The fit converges to a high
but poorly-constrained calcium 
abundance of Ca $>$ 4.7 $\times$ solar
and Fe = 1.8 [0.8 - 3.1; 1$\sigma$] $\times$ solar.

\noindent \underline{ G75.84$+$0.40}
A  diffuse image for the region near G75.84$+$0.40 (Fig. 9-right) 
was constructed using the same procedure as above.
The residual emission after expunging point sources is somewhat
more localized near the radio H II peak positions than for G75.77$+$0.34.
For comparison, diffuse spectra were extracted from the circular 
region and the triangular region shown in Fig. 9-right. 
The circular region covers most of the diffuse emission but
7 X-ray sources were detected inside this region. Their emission
was excluded from the diffuse spectral extraction. The
smaller triangular region does not enclose any detected X-ray sources
but contains fewer net diffuse counts. After subtracting nominal
background the r = 14$''$ circular region contains 131 net counts
(0.2 - 8 keV) and  the smaller triangular region contains
54 net counts.

Unlike the diffuse emission spectrum of G75.77$+$0.34,
that of G75.84$+$0.40 shows no hard emission above 
5 keV. There are no conspicuous emission lines but a
line-like feature is visible in lightly-binned spectra
at 3.9 keV may be Ca XIX (log T$_{max,{\rm Ca XIX}}$ = 7.5 K).
Fits of the spectra for the two different regions with a 
1T $apec$ thermal plasma model give similar  results
and the fit for the higher signal-to-noise spectrum
extracted from the larger circular region is summarized in Table 6.
The absorption from the 1T $apec$ fit converts to 
A$_{\rm V}$ = 22.1 [16.8 - 33.2] mag.
The temperature determined from thermal models (kT = 2.2 keV) is
lower than obtained above for the G75.77$+$0.34 diffuse emission
but the N$_{\rm H}$ values are similar.

Since there are no conspicuous lines in the G75.84$+$0.40 diffuse
spectrum with the possible exception of Ca XIX, we also fitted
the spectrum with a power-law model (Table 6). It yields a slightly better
fit as gauged by $\chi^2$. However, the best-fit absorption
N$_{\rm H}$ = 6.2 $\times$ 10$^{22}$ cm$^{-2}$ is suspiciously large,
being a factor of $\approx$3 larger than the values determined for 
point sources in the G75.84$+$0.40 region (Table 4; Sec. 4.3.1). 
The photon power-law index $\Gamma_{ph}$ = 3.9 [3.0 - 4.9; 1$\sigma$] 
equates to the
energy power-law index $\Gamma_{E}$ according to 
$\Gamma_{E}$ = $\Gamma_{ph}$ $-$ 1 = 2.9 [2.0 - 3.9].
That is, the observed flux scales with energy according to
F$_{x}$ $\propto$ E$^{-2.9 \pm 0.9}$. This is a rather steep
falloff as compared to that found for the diffuse power-law
spectrum of RCW 38 ($\Gamma_{E}$ = 1.59 [1.32 - 2.77]; 
Wolk et al. 2002), but is nearly identical to that 
determined for the diffuse emission of Westerlund 1
($\Gamma_{E}$ = 2.7 $\pm$ 0.2; Muno et al. 2006).

\input{table6.tex}

\section{Discussion}

\subsection{Where are the Classical T Tauri Stars?}

At the young age of Berkeley 87 ($\sim$3 - 5 Myr), a population of
low-mass pre-main sequence stars (T Tauri stars) should be present. 
Classical T Tauri stars (cTTS) are still surrounded by disks and
have spectral energy distributions showing  excess near-IR 
emission originating in warm circumstellar material.
In addition, weak-lined T Tauri stars (wTTS) with  
little or no circumstellar disk material may be present but 
are not expected to have significant near-IR excesses.
Both cTTS and wTTS emit strongly in X-rays  and can undergo 
significant variability including bright X-ray flares.
We have thus searched for near-IR counterparts in the 2MASS and
UKIDSS GPS catalogs within 1$''$ of the X-ray positions of all 
{\em Chandra} ON 2 sources. Color-color diagrams were constructed 
to determine which {\em Chandra} counterparts have near-IR excesses.
We did not compute mid-IR colors since {\em Spitzer} IRAC coverage
for ON 2 is incomplete and hampered by high IR surface 
brightness in the northern part near BC Cyg.

We found 2MASS counterparts for less than half of the {\em Chandra} sources
and many of these lack photometry in all three JHK$_{s}$ bands, so colors
could not be computed. The more sensitive UGPS survey yielded
counterparts for 78\% of the {\em Chandra} sources, and about
80\% of these have complete JHK photometry. The UGPS 90\%  completeness
detection limits in uncrowded regions are J = 19.5, H = 18.75, and K = 18.0,
but limits can be $\approx$1 magnitude less in crowded 
fields  (Lucas et al. 2008). These limits are sufficient to
detect TTS in ON 2. For comparison, we note that X-ray detected
YSOs in the {\em Chandra} COUP ONC survey span a wide range of 
near-IR magnitudes, but typical values are 
J, K$_{s}$ $\approx$ 11 - 13 mag (Getman et al. 2005).
Scaling these magnitudes from the ONC distance of 0.45 kpc
to Berkeley 87 (1.75 kpc) gives J, K$_{s}$ $\approx$ 13.9 - 15.9.
At the assumed distance of G75.77$+$0.34 (3.5 kpc), the scaled values
are J, K$_{s}$ $\approx$ 15.4 - 17.4.  

Figure 10-top is a near-IR color-color diagram for 
those UGPS sources in ON 2 that were detected at 
J,H, and K and are associated with a {\em Chandra} source.
As the figure shows, most UGPS near-IR counterparts 
have colors consistent with normally-reddened stars
and a normal reddening law for this region is 
anticipated based on the findings of TF82.
Of more interest here are the 20 UGPS sources lying
to the right (redward) of the A0V line whose properties are
listed in Table 7. The spatial positions of these YSO
candidates are plotted in Figure 11 along with positions 
of some X-ray detected Berkeley 87 stars.

Of the 20 sources lying redward of the A0V line in
Figure 10-top, thirteen lie on or above the line marking the locus of
{\em unreddened} cTTS (Meyer et al. 1997).  These objects are
identified in Table 7 and are candidate cTTS, 
but may also include some extragalactic objects. 
Four of these thirteen have a high probability of 
X-ray variability, typical of magnetically-active 
late-type pre-main sequence stars.
Two of these four variable sources appear to be
heavily-reddened based on UGPS colors (UGPS J202134.90$+$372442.0
and J202142.66$+$373031.4). Their high median photon energies
could also be due to high absorption.


Figure 10-bottom is a UGPS near-IR color-color diagram
showing colors of the 72 UGPS sources with detections in
all bands lying within a 1$'$$\times$1$'$ region centered
on G75.77$+$0.34 (see Fig. 4-top for the {\em Spitzer} view). 
This figure includes {\em all}  UGPS sources
in the region with complete near-IR  photometry, and those 
detected  by {\em Chandra} are circled in red. All of
the {\em Chandra} detections in Table 2 are included
except two sources numbered  7 and 15 in the first column, 
whose UGPS counterparts lack complete photometry.
There are 17 UGPS sources
lying redward of the A0V line but only one of these
is detected by {\em Chandra} (source 10 in Table 2).
Its near-IR UGPS photometry may be 
affected by a second closely-spaced  source.

Even though Figure 10-bottom reveals several near-IR
excess sources in the G75.77$+$0.34 region, 
they are not detected by {\em Chandra} with
one exception, as noted above. It is thus clear that 
the X-ray sources near G75.77$+$0.34 listed in Table 2 
do not have near-IR colors consistent with that
expected for cTTS. Instead, the X-ray detected sources
are a combination of reddened field stars,
extragalactic objects, and YSOs with colors
different from reddened cTTS. The latter group could 
include wTTS without significant near-IR excesses, 
which cannot be distinguished from normally-reddened 
field stars using near-IR colors. In addition, some
massive YSOs may be present in Table 2 such as
the heavily-obscured source 
lying inside the radio position error box 
of G75.77$+$0.34B (CXO J202140.74$+$372619.6; Sec. 3.2).

As can be seen in Figure 11, the optically-revealed
Berkeley 87 stars  lie  to the south of G75.77$+$0.34.  
They are viewed through moderate extinction 
A$_{\rm V}$ $\approx$ 4.8 mag (TF82). 
In contrast, the near-IR excess stars lie northward
and probably trace a younger and more heavily-obscured
population than  the visible Berkeley 87 stars. 
Most of the near-IR excess sources lack optical counterparts and
their distances are not known. They could lie further
away at the H II region distances ($\gtsimeq$3.5 kpc) and
be unrelated to the optically-identified Berkeley 87 stars 
to the south.

To summarize the above, we find only about a dozen
near-IR excess counterparts to {\em Chandra} sources 
in ON 2 whose colors are consistent with cTTS.
And, quite remarkably, only one of the {\em Chandra}
sources in the crowded region near G75.77$+$0.34
listed in Table 2 has a near-IR excess counterpart.
The low number of {\em Chandra} detections of near-IR
excess sources in the G75.77$+$0.34 region is probably
due to insufficient sensitivity. The 5-count ACIS-I detection
threshold in this region is log L$_{x}$ = 30.77 ergs s$^{-1}$
at d = 3.5 kpc (Sec. 4.3.1). This limit lies at the high 
end of the mass range for TTS (M$_{*}$ $\gtsimeq$ 1.5 M$_{\odot}$)
and very few TTS were  detected in the COUP ONC or Taurus Molecular Cloud X-ray surveys  
with log L$_{x}$ $\geq$ 30.77 (Preibisch et al. 2005; Telleschi et al. 2007).

\input{table7.tex}

\subsection{X-rays from WR 142 }

X-ray emission from  WR 142 has now been clearly detected in 
two {\em Chandra} observations as discussed herein and in
Sokal et al. (2010), and by {\em XMM-Newton} (Oskinova et al. 2009). 
WR 142 is an important benchmark since it provides insight into 
the X-ray properties of highly-evolved WO stars which are
supernova progenitors  and extremely rare in the Galaxy 
(Tramper et al. 2015).

The X-ray emission of single massive WR and OB stars without known
companions is usually attributed to radiative shocks that form in their
strong winds via  line-driven instabilities (e.g. Lucy \& White 1980;
Owocki et al. 1988; Gayley \& Owocki 1995; Feldmeier et al. 1997). 
In addition, X-ray
emission is predicted to occur as a result of wind-wind collision
in the region between massive binaries such as WR$+$O systems
(Prilutskii \& Usov 1976; Luo et al. 1990; Usov 1992).
For single stars, relatively low  temperatures kT $\ltsimeq$ 1 keV
are predicted for X-rays formed in radiative wind shocks.
This prediction is at odds with the high X-ray temperatures 
of several keV determined for WR 142 using thermal models (Table 5),
given that WR 142 has so far shown no evidence for a companion.
A similar situation exists for putatively single nitrogen-type 
WN stars (Skinner et al. 2010; 2012). It is thus apparent that
even though radiative wind shock models have had some success 
explaining soft X-ray emission from single O-type stars, their
ability to explain the harder X-ray emission from single WR
stars with higher mass-loss rates and wind speeds remains
to be shown.

Several possible X-ray emission mechanisms for WR 142 were
discussed by Sokal et al. (2010) including both thermal and
nonthermal (e.g. inverse Compton scattering) processes.
That discussion remains largely unchanged in light of the 
new {\em Chandra} detection of WR 142 in 2016 and we refer
the reader to Sokal et al. (2010) for details.
But we summarize below a few
new findings for WR 142 that emerge from the second
{\em Chandra} observation.

First, the observed broad-band (0.2-8 keV) X-ray fluxes
are comparable in the two {\em Chandra} observations 
taken  $\approx$7.5 years apart. Flux measurements
in the individual ACIS-I spectra are model-dependent
and subject to low-count statistics, especially in
the 2016 observation where WR 142 was captured far
off-axis. Measurement uncertainties allow for 
a difference in the observed flux of up to $\approx$30\% between
the two observations. But given the difficulty in obtaining
accurate flux measurements in the low-count spectra
it is quite possible that little or no change
in X-ray flux occurred. Further X-ray monitoring would
be needed to determine if any significant flux
variability is present, but the case for flux variability
based on existing {\em Chandra} data is not compelling.

Second, the recent {\em Gaia} DR2 distance for WR 142
provides a more accurate X-ray luminosity determination
log L$_{x}$ = 31.15$\pm$0.15 ergs s$^{-1}$ than in 
previous studies, where the uncertainties reflect the
spread in values obtained from different spectral models (Table 5).
This value is at the low end of the range for X-ray detected
single WR stars (Fig. 4 of Skinner et al. 2012). Similarly,
the ratio log (L$_{x}$/L$_{bol}$) = $-$7.90$\pm$0.15                                                                               
(Table 5) is low compared to other X-ray detected WR stars and O-type
stars (Fig. 5 of Sokal et al. 2010). Interestingly, the 
subclass of single carbon-type WC stars remains undetected
in X-rays so far with upper limits on L$_{x}$  that are in
some cases below the value quoted above for WR 142.
Although a L$_{x}$ $\propto$ L$_{bol}$ correlation
exists for O-type stars, such a correlation has not
yet been found for WR stars. Other properties besides
L$_{bol}$ such as mass-loss parameters and abundances
must be taken into account in models of the X-ray emission
detected from WN and WO stars, or lack thereof in  
single WC stars.

Finally, the 2016 spectrum
shows indications of spectral lines indicative
of thermal X-ray emission. Although no clearly discernible
lines were seen in the 2009 {\em Chandra} spectrum,
spectral fits showed some improvement with models 
that included a Gaussian Fe line at 6.67 keV and
a weak spectral peak near 4 keV may be present 
(Sokal et al. 2010).

A thermal interpretation for the WR 142 X-ray emission 
is favored given that thermal emission at several keV
is the norm for X-ray detected WR stars.
X-ray emission models are thus confronted with
explaining very hot thermal plasma in WR 142,
including possible Fe K line emission (6.67 keV).
The Fe K  emission-line complex is due to  Fe XXV lines 
which form at very high characteristic temperatures
near T$_{max, {\rm Fe XXV}}$ $\sim$ 60 MK. 
Fe K emission has been detected in
a few other putatively single WR stars 
including WR 134 (WN6) and WR 20b (WN6ha),
as shown in Skinner et al. (2010).

The intriguing question that theoretical models must 
address is how apparently single WR stars, which are not 
known to be  colliding wind binaries, can produce
X-ray plasma at such high temperatures. Because of
the high mass loss rates and wind speeds of WR stars,
their winds provide a huge reservoir of kinetic energy  
that can be tapped to produce shock-induced X-rays. 
Using the WR 142 mass-loss
parameters adopted in Sokal et al. (2010), namely
$\dot{\rm M}$ = 1.7 $\times$  10$^{-5}$ M$_{\odot}$ yr$^{-1}$
(unclumped) and $v_{\infty}$ =  5500 km s$^{-1}$,
the wind kinetic luminosity is L$_{wind}$ =
(1/2)$\dot{\rm M}$$v_{\infty}^2$ = 1.6 $\times$ 10$^{38}$ 
ergs s$^{-1}$. This gives log (L$_{x}$/L$_{wind}$) $\approx$ $-$6.9.
Thus, there is sufficient energy in the wind to account for
the X-ray luminosity even if the conversion efficiency is low,
and the very high wind speed of WR 142 is capable of producing
shock temperatures of several keV (Sokal et al. 2010).
In addition to further refinement of wind shock models for
single WR stars, the possible role of magnetic fields in
producing hot X-ray plasma deserves further study.
But in the absence of magnetic field detections for 
WR stars, such models cannot yet be rigorously tested. 

\subsection{Comments on Diffuse X-ray Emission in ON 2}

The apparent diffuse X-ray emission shown in Figure 9
is localized in the immediate vicinity of 
G75.77$+$0.34 and G75.84$+$0.40. Analysis of
the diffuse spectra (Sec. 3.8) suggests high-temperature
thermal emission but nonthermal emission is not
ruled out for G75.84$+$0.40. Extended thermal emission could arise from 
a population of undetected faint point sources or
from truly diffuse emission. Given that the diffuse emission
is localized near the H II regions, shocked winds of the 
embedded massive stars may be involved.

\subsubsection{Faint Point Sources Masquerading as Diffuse X-ray Emission?}

\noindent \underline{G75.77$+$0.34}~

As discussed above, thermal models of the diffuse spectrum of the
region near G75.77$+$0.34 give an absorption
N$_{\rm H}$ = 3.6 [2.8 - 4.9; 1$\sigma$] $\times$ 10$^{22}$ cm$^{-2}$
and kT = 5.4 [3.3 - 9.1; 1$\sigma$] keV. The {\em Chandra}
unabsorbed flux detection limit (ObsId 9914; 5 count threshold) for a generic 
low-mass pre-main sequence star (TTS) with a typical thermal plasma 
spectrum at kT $\approx$ 3 keV viewed through the above absorption is 
F$_{x,unabs}$(0.3-8 keV) = 4.0 $\times$ 10$^{-15}$
ergs cm$^{-2}$ s$^{-1}$, equivalent to  
log L$_{x}$ = 30.77 ergs s$^{-1}$ at d = 3.5 kpc.
Our 1T $apec$ model fits of the G75.77$+$0.34 diffuse spectrum
give an unabsorbed net flux 
F$_{x,unabs}$(0.3-8 keV) = 1.14 $\times$ 10$^{-13}$ ergs cm$^{-2}$ s$^{-1}$,
or log L$_{x,diffuse}$ = 32.22 ergs s$^{-1}$.
It would thus require at least 28 undetected faint sources ($<$5 counts) 
within the diffuse spectrum extraction ellipse (area = 867 arcsec$^2$) to 
account for the observed luminosity, as compared to only one point source 
actually detected within the  ellipse (CXO J202140.75$+$372533.4; 60 net cts).

The above estimate of 28 undetected sources is a strict lower limit based
on the {\em Chandra} 5-count detection threshold and is undoubtedly an underestimate.
It assumes an average luminosity per star of log L$_{x}$ = 30.77 ergs s$^{-1}$
which is at the high end of X-ray luminosities observed for young low-mass
stars in the deep {\em Chandra} Orion Nebula Cluster survey (COUP),
as shown in Figure 3 of Feigelson et al. (2005). A more representative value
would be the median (or mean) luminosity of the 559  heavily-obscured young stars 
detected in COUP. These stars have a mean absorption 
N$_{H,mean}$ = 3.2 $\times$ 10$^{22}$ cm$^{-2}$  
(similar to that of the
G75.77$+$0.34 diffuse emission spectrum), median log L$_{x,med}$ = 29.85 ergs s$^{-1}$,
and mean log L$_{x,mean}$ = 30.43 ergs s$^{-1}$ (Table 1 of Feigelson et al. 2005).
Adopting the median value, about 230 heavily-obscured young stars would be needed to 
account for the integrated luminosity in the G75.77$+$0.34 diffuse extraction ellipse,
and about 60 sources are required if  L$_{x,mean}$ is adopted.
A search of the UKIDSS GPS data base reveals 81 sources
within the diffuse extraction ellipse, excluding spurious
sources classified as noise. Based on more reliable UGPS source class
distributions in less-crowded fields away from H II 
regions in ON 2S, we estimate about 20\% of these 
are galaxies or probable galaxies. This leaves 
$\approx$65/81 UGPS sources within the ellipse that
are probable stars. Almost all of these would need
to have X-ray luminosities at or above the ONC
L$_{x,mean}$ value to account for the diffuse
X-ray luminosity, which is rather unlikely.

Fits of the spectra of four {\em Chandra} sources
lying just outside the diffuse extraction ellipse and
the source inside it 
with 1T $apec$ models give a mean (median) 
N$_{\rm H}$ = 2.8 (3.0) $\times$ 10$^{22}$ cm$^{-2}$
and kT = 3.8 (3.9) keV. These values are  less
than obtained for the diffuse spectrum but are within
the 1$\sigma$ confidence ranges of the diffuse fit.
However, the spectra of the five representative sources
do not show the features at 4.1, 5.2, and 6.7 keV that
are visible in the diffuse spectrum. A simulated 1T $apec$
point source X-ray spectrum using the median N$_{\rm H}$ and 
kT values above has maximum counts in the 1.5 - 2 keV range
whereas the diffuse spectrum in Figure 9 peaks at a higher
energy above 2 keV. Simulated spectra with artificially high count
rates  do show the 6.7 keV line but not the 4.1 or 5.2 keV features.

\noindent \underline{G75.84$+$0.40}~ 
Using the higher signal-to-noise diffuse spectrum from
the r = 14$''$ circular extraction region (area = 616 arcsec$^{2}$)
shown in Figure 9-top gives an unabsorbed net diffuse flux
F$_{x}$(0.3-8 keV) = 1.81 $\times$ 10$^{-13}$ ergs cm$^{-2}$ s$^{-1}$.
Assuming d = 3.5 kpc as adopted above for G75.77$+$0.35 gives an
intrinisic diffuse X-ray luminosity log L$_{x,diffuse}$ = 32.42 ergs s$^{-1}$.
The {\em Chandra} 5-count detection threshold in ObsId 18083
is log Lx(0.3-8 keV) = 30.82 ergs s$^{-1}$ (d = 3.5 kpc)
for a generic TTS viewed through the diffuse spectrum $apec$ model
absorption N$_{\rm H}$ = 4.2 [3.2 - 6.3] $\times$ 10$^{22}$ cm$^{-2}$. 
Thus, a  minimum of $\approx$40 faint ($<$5 counts) X-ray sources 
would be needed to account for the integrated diffuse X-ray luminosity 
within the G75.84$+$0.40 extraction region. By comparison, 
only 7 X-ray sources were detected inside the extraction region.
 
As already noted for G75.77$+$0.34, the minimum value of 40 faint
sources computed using  the {\em Chandra} detection limit is likely
an underestimate. The COUP L$_{x,med}$ value given above yields a more 
realistic estimate of  $\approx$370 X-ray
sources within the diffuse extraction circle, and the COUP
L$_{x,mean}$ value gives $\approx$98 sources.
No reliable estimate of the number of stars actually present
can be obtained from the UKIDSS GPS or {\em Spitzer} source
catalogs because of the adverse effects of high IR surface
brightness in this region due to BC Cyg. 
Only four UKIDSS GPS sources are cataloged within the extraction circle. 
Scaling down the estimate obtained above for G75.77$+$0.34
to the slightly smaller diffuse extraction region used for
G75.84$+$0.40 suggests  $\approx$46 near-IR sources classified
as stars should be present within the r = 14$''$ extraction circle,
well below the number needed to account for the diffuse X-ray luminosity.

Thermal model fits of the X-ray spectra of five sources lying 
inside or near the circular diffuse extraction region give a 
mean (median) N$_{\rm H}$ = 2.5 (2.6) $\times$ 10$^{22}$ cm$^{-2}$,
about a factor of two less than that measured in the diffuse
spectrum. These five sources require high plasma temperatures
kT $\geq$ 7 keV if thermal emission is assumed  but the signal-to-noise
in their spectra is not sufficient to distinguish between thermal and 
nonthermal emission.

In summary, faint ($<$5 counts) undetected X-ray sources
could contribute to some of the diffuse emission near 
the H II regions discussed above, but attributing all of the 
diffuse emission to faint point sources seems very unlikely
based on estimates of the required number of faint X-ray sources
and comparisons of the diffuse X-ray spectra with detected point 
sources. We thus conclude that some true diffusion emission is 
indeed present.

\newpage

\subsubsection{Wind Shock Processes}

The embedded stars in the H II regions are either  O
or early B-type stars which are expected to  have 
powerful supersonic winds. These winds can interact with
each other and with surrounding material to produce
hot shocked plasma at X-ray emitting temperatures.

The temperature of the hottest shocked plasma is
predicted to be 
T$_{shock}$ = 15.5$v_{w,1000}^2$ MK where
$v_{w,1000}$ is the wind speed in units
of 1000 km s$^{-1}$ (Cant\'{o} et al. 2000).
Fits of the diffuse spectra with thermal models (Table 6)
give T = 63 MK (kT = 5.4 keV)  for G75.77$+$0.34
and T = 25 MK (kT = 2.2 keV)  for G75.84$+$0.40.
To achieve such temperatures via wind shocks
the required wind speeds are at least
$v_{w}$ $\approx$2000 km s$^{-1}$ (G75.77$+$0.34)
and $\approx$1300 km s$^{-1}$ (G75.84$+$0.40).
The higher speed for G75.77$+$0.34 is compatible
with the wind of a mid or late OV  star and the
lower speed is characteristic of early B-type
stars (Lamers \& Leitherer 1993; Kudritzki \& Puls 2000;  
Macfarlane, Cohen, \& Wang 1994).

Can the wind kinetic energy of a mid to late OV star
or a $\approx$B0V star account for the diffuse
X-ray luminosities? To make this comparison we 
estimate the wind kinetic luminosity
L$_{w}$ = (1/2)$\dot{M}$$v_{w}^2$ =
3.17 $\times$10$^{35}$$\dot{M}_{-6}$$v_{w,1000}^2$ ergs s$^{-1}$
where $\dot{M}_{-6}$ is the stellar mass loss rate
in units of 10$^{-6}$ M$_{\odot}$ yr$^{-1}$
and $v_{w,1000}$ is the wind speed in units of 1000 km s$^{-1}$.
For the diffuse emission near G75.77$+$0.34 we obtain
log L$_{x,diffuse-total}$ = 32.33 ergs s$^{-1}$ after 
adjusting the value in Table 6 upward by 30\% to account
for the diffuse emission that was not captured in the 
extraction ellipse (Fig. 9-left). The above luminosity
could be produced by a single $\approx$B0V star  
($\dot{M}$ $\sim$ 10$^{-8}$ M$_{\odot}$ yr$^{-1}$,
$v_{w}$ $\sim$ 1500 km s$^{-1}$, 
log L$_{w}$($\approx$B0V) $\sim$ 33.8 ergs s$^{-1}$)
or a $\approx$O7V - O9V star
($\dot{M}$ $\sim$ 10$^{-7}$ M$_{\odot}$ yr$^{-1}$,
$v_{w}$ $\sim$ 2000 km s$^{-1}$,
log L$_{w}$($\approx$O7-9V) $\sim$ 35.1 ergs s$^{-1}$).
These spectral types are consistent with the range
derived for G75.77$+$0.34B (Sec. 3.2).
A similar conclusion holds for the diffuse emission near
G75.84$+$0.40B.

\subsubsection{Diffuse X-ray Plasma Properties}

Using the diffuse X-ray spectral fits for G75.77$+$0.34 in Table 6
and following the methodology used by Townsley et al. (2003) for
diffuse emission in the Rosette Nebula and M17, we can derive
basic physical properties of the diffuse plasma. The electron
density $n_{e}$ in the X-ray emitting region is given by
L$_{x}$ = $\Lambda$(T)$n_{e}^2$$\eta$V, where 
$\Lambda$(T) is the temperature-dependent volumetric cooling
rate, V is the volume of the X-ray plasma, and the free parameter
$\eta$ $<$ 1 is the X-ray plasma volume filling factor. We assume an ellipsoidal
emitting volume with angular semi-axes 
23$''$ $\times$ 12$''$ $\times$ 12$''$ corresponding to 
the diffuse extraction ellipse in Figure 9-top-left.
For the diffuse X-ray luminosity log L$_{x,diffuse}$ = 32.22 ergs s$^{-1}$ 
(Table 6; d = 3.5 kpc) measured within the extraction ellipse, the above gives
V = 1.49 $\times$ 10$^{54}$ cm$^{3}$
and $n_{e}$ = 2.4/$\sqrt{\eta}$ cm$^{-3}$. Here, we have
used $\Lambda$(T = 63 MK) $\approx$ 2 $\times$ 10$^{-23}$ erg cm$^{3}$ s$^{-1}$
at the X-ray temperature kT = 5.4 keV based on the approximation 
given in van den Oord \& Mewe (1989).  The mass of the hot X-ray plasma 
is M$_{X}$ = $\mu$$n_{e}$m$_{\rm H}$V 
$\approx$ 1.85 $\times$ 10$^{-3}$/$\sqrt{\eta}$ M$_{\odot}$,
where $\mu$ = 0.62 (amu) is the mean  weight per particle for
fully-ionized cosmic abundance plasma and m$_{\rm H}$ is the proton mass.

The radiative cooling time of the X-ray plasma is
$\tau_{cool}$ = 3kT/$n_{e}$$\Lambda$(T) $\approx$ 17$\sqrt{\eta}$ Myr.
This is much longer than the dynamical equilibration timescale 
D/$v_{s}$ $\sim$ 700 yr, where D is the diameter of the emitting region and
$v_{s}$ $\approx$ 794 km s$^{-1}$ is the sound speed in the hot X-ray plasma.
The time needed to refill the 
emitting volume is $\tau_{refill}$ $\sim$ 18,500/$\sqrt{\eta}$ y, 
assuming mass replenishment from a O-type star with a typical
mass loss rate $\dot{M}$ $\sim$ 10$^{-7}$ M$_{\odot}$ yr$^{-1}$.
Combining the above with the approximation 
$\eta$ $\sim$ M$_{X}$/($\dot{M}$$\tau_{cool}$) yields 
the estimate $\eta$ $\sim$ 0.033, assuming as above
$\dot{M}$ $\sim$ 10$^{-7}$ M$_{\odot}$ yr$^{-1}$.
Thus, the hot X-ray plasma need only comprise a
few percent of the total gas volume, the dominant
component being  cooler wind gas.

For the above value of $\eta$, the cooling time is
$\tau_{cool}$ $\sim$ 3 Myr. This is greater than
the estimated age of an embedded O-type star,
$\tau_{*}$ $\ltsimeq$ 2 Myr. In this case,
the above
expression for $\eta$ should be replaced by
$\eta$ $\sim$ M$_{X}$/($\dot{M}$$\tau_{*}$).
Taking $\tau_{*}$ $\sim$ 2 Myr then gives 
$\eta$ $\sim$ 0.05, $\tau_{cool}$ $\sim$ 3.8 Myr
and $\tau_{refill}$ $\sim$ 0.1 Myr.
Thus, little if any cooling of 
the hot plasma has occurred. In addition,
the hot plasma is replenished  much faster than it 
cools and the amount 
of hot diffuse plasma is increasing with time.

\subsubsection{High-Temperature Emission Lines in Diffuse X-ray Spectra}

The presence of high-temperature Ca XX and Fe K lines in the 
diffuse spectrum of G75.77$+$0.34 is noteworthy since some
diffuse emission models predict such lines. In the model of
Dorland \& Montmerle (1987), wind energy from massive 
stars in H II regions can dissipate energy near the 
shock and produce a hot X-ray tail from a population of
non-Maxwellian electrons. Diffuse X-ray temperatures
scale with the wind speed and values of kT $\approx$5 keV
or higher can be achieved and high-temperature lines such as Fe K
can be produced.

Their theory predicts the value of a flux-limit 
factor $\zeta$ which depends on the wind speed
and temperature of the hot X-ray plasma.
For specific values of $\zeta$ and $v_{w}$
their model predicts the ratio 
$R$ = L$_{x,diffuse}$(0.2 - 4 keV)/L$_{w}$.
In the case of G75.77$+$0.34 (kT $\approx$ 5 keV)
their graphical results predict
$R$ $\approx$ 7 $\times$ 10$^{-4}$
for a ``slow'' wind ($v_{w}$ = 1000 km s$^{-1}$)
and $R$ $\approx$ (2 - 3) $\times$ 10$^{-4}$
for ``fast'' winds ($v_{w}$ = 2000 - 3000 km s$^{-1}$). 
If the diffuse flux  measurement of G75.77$+$0.34
is restricted to the 0.2-4 keV range used by
Dorland \& Montmerle and adjusted upward by
30\% to account for diffuse emission lying outside 
the extraction ellipse (Fig. 8-top), then 
log L$_{x,diffuse-total}$(0.2 - 4 keV) = 32.17 ergs s$^{-1}$.
The predicted wind
luminosities are log L$_{w}$ = 35.3 ergs s$^{-1}$
(slow wind) and log L$_{w}$ = 35.7 - 35.9 ergs s$^{-1}$
(fast wind). These values are achievable by  single
$\approx$O6V - O7V stars having fast winds
($\dot{M}$ $\approx$ few $\times$ 10$^{-7}$ M$_{\odot}$ yr$^{-1}$,
$v_{w}$ $\approx$ 2500 km s$^{-1}$). These spectral
types are slightly earlier than the late OV or 
early BV range  obtained from near-IR and
radio estimates (Sec. 3.2). However, the estimate
based on the Dorland \& Montmerle model assumes
a single exciting star whereas multiple sources
may be contributing in the crowded G75.77$+$0.34 region.
In addition, mass-loss rates, wind speeds, and 
L$_{w}$ of the exciting
stars in G75.77$+$0.34 are not well-determined.

Finally, we call attention to the remarkable 
similarity between the 2016 X-ray spectrum of 
WR 142 (Fig. 7) and the diffuse spectrum
of G75.77$+$0.34 (Fig. 9-top). Both spectra
show the Fe K line at 6.7 keV and the feature
at 4.1 keV which we attribute to Ca XX.
The unusual feature at 5.2 keV visible in
the G75.77$+$0.34 spectrum is not obviously
detected in WR 142 but some emission slightly
above 5 keV is seen in WR 142, even at the
low signal-to-noise of its spectrum.
In addition, thermal models give similar
plasma temperatures kT $\approx$ 5 keV for
the G75.77$+$0.34 diffuse emission and 
WR 142 (Tables 5 and 6). We note that
thermal fits of the hard component of
the WR-rich cluster Westerlund 1 give
similar temperatures, but no emission
lines such as Fe K were detected in
its hard component (Muno et al. 2006).

The remarkable similarity between the 
diffuse spectrum of G75.77$+$0.34
and the WR 142 spectrum raises the interesting  
possibility that the same physical 
process underlies both spectra, 
which could realistically be shocked winds.
However, WR 142 is detected as a
(faint) point source, so its X-ray emission
is not diffuse at the sensitivity of
our {\em Chandra} observations.
In future work, a deep X-ray observation of WR 142
to search for diffuse  emission in
its immediate vicinity might be worthwhile.
Such an observation is further motivated by
the presence of diffuse C IV (5801 - 5812 \AA)
emission near WR 142, as seen in the 
optical study of Polcaro et al. (1991).
They attribute the C IV emission to 
high-temperature shock fronts surrounding
WR 142 and speculate that such shocks
could explain faint diffuse X-ray emission 
in the region reported on the basis of
{\em EXOSAT} data (Warwick et al. 1988).

\section{Conclusions}

Two sensitive {\em Chandra} observations of the Onsala 2 
star-forming region   have provided new X-ray images of
unprecedented sensitivity and spatial resolution, along
with undispersed X-ray spectra of high-interest sources. 
The main conclusions are as follows.

\begin{enumerate}

\item 
More than 300
X-ray sources were detected in ON 2 including
optically-identified OB stars in the Berkeley 87
cluster and the rare  oxygen-type Wolf-Rayet star 
WR 142. The most conspicuous 
emission occurs near the  compact H II regions G75.77$+$0.34 and
G75.84$+$0.40 where dense groupings of X-ray point sources
are found along with diffuse emission. 
Previous radio parallax measurements compared with
new {\em GAIA} DR2 stellar parallaxes reveal that these H II regions
are more distant than Berkeley 87, and in all likelihood not
physically associated with the optically-revealed cluster.

\item An X-ray point source with a near-IR counterpart
is detected at H II region radio continuum peak G75.77$+$0.34B. 
The inferred spectral type based on near-IR and radio 
data is late O or early B, but the derived spectral type
is distance-dependent.

\item Two X-ray sources separated by 2$''$.6 are detected
near the H II region radio continuum peak G75.84$+$0.40B.
The presence of multiple X-ray and 
IR sources near G75.84$+$0.40B suggests that more than
one embedded young OB star may be exciting the H II region.

\item Diffuse X-ray emission is present near 
G75.77$+$0.34 and G75.84$+$0.40. The diffuse spectrum
of G75.77$+$0.34 shows high-temperature emission lines 
including Fe K (6.67 keV) indicative of hot thermal plasma.
It is unlikely that a population of faint X-ray sources
can account for the diffuse emission. 
Shocked winds from the embedded massive stars
offer a plausible explanation.

\item Only about a dozen X-ray sources were identified in ON 2 
with near-IR excess counterparts having colors typical of classical 
T Tauri stars. Remarkably, only one {\em Chandra}
source in the crowded region near G75.77$+$0.34 has a near-IR
excess. Although a population of cTTS may be present in
the G75.77$+$0.34 region, such a population has largely escaped 
X-ray detection, perhaps because of insufficient sensitivity. 
At the assumed distance of the 
H II region ($\sim$3.5 kpc), {\em Chandra} is
capable of detecting only the most X-ray luminous
and massive cTTS. Subsolar mass cTTS could 
have easily evaded detection.

\item The X-ray emission of WR 142 is heavily-absorbed,
probably by its metal-rich wind, and its spectrum 
acquired in 2016 reveals
faint emission lines indicative of very hot thermal
plasma (kT $\approx$ 5 keV). Such hot plasma is not predicted
by X-ray wind shock models of massive single stars and
further development of X-ray emission models for single
WR stars is needed. The X-ray luminosity of WR 142
determined from its {\em Gaia} DR2 distance 
(log L$_{x}$ = 31.15$\pm$0.15 ergs s$^{-1}$)
is at the low end of the range found for WR stars.

\item The brightest X-ray source detected by {\em Chandra}
is an unusual object that is revealed in the IR. The absence
of an optical counterpart and high absorption in its X-ray
spectrum argue against a foreground object. 
Its IR and X-ray properties are suggestive of a
relatively  massive class I protostar or a heavily-reddened 
class II source.

\end{enumerate}

\acknowledgments

Support for this work was provided by the National Aeronautics Space Administration (NASA) through 
{\em Chandra} award number GO6-17133X issued by the {\em Chandra} X-ray Center, which is operated by
the Smithsonian Astrophysical Observatory (SAO) for and on behalf of NASA. 
This work has utilized data obtained in the Two Micron All-Sky Survey (2MASS) and
the UKIRT {\em Deep Sky Survey}. This work is based in part on archival data obtained
with the {\em Spitzer Space Telescope}, operated by the Jet Propulsion Laboratory, California 
Institute of Technology under contract with NASA.


\vspace{5mm}
\facilities{Chandra(ACIS-I)}

\clearpage


\input{f1.tex}

\input{f2.tex}

\clearpage

\input{f3.tex}

\clearpage

\input{f4.tex}

\clearpage

\input{f5.tex}

\clearpage

\input{f6.tex}

\clearpage

\input{f7.tex}

\clearpage

\input{f8.tex}

\clearpage

\input{f9.tex}

\clearpage

\input{f10.tex}

\clearpage

\input{f11.tex}

\clearpage

\end{document}

%% file: table1.tex
\begin{deluxetable}{lll}
\tabletypesize{\small}
\tablewidth{0pt}
\tablecaption{Chandra X-ray Observations of Onsala 2  }
\tablehead{
\colhead{Parameter} &
\colhead{} \\
}
\startdata
Region               & ON2S                       & ON2N     \\
ObsId                & 9914\tablenotemark{a}      & 18083\tablenotemark{b}    \\
Start Date/Time (TT) & 2 Feb. 2009/02:41:24       & 13 Aug. 2016/09:29:27     \\
Stop  Date/Time (TT) & 2 Feb. 2009/22:52:55       & 14 Aug. 2016/05:31:06       \\
Instrument           & ACIS-I\tablenotemark{c}    & ACIS-I\tablenotemark{c}  \\                                                           
Livetime (s)         & 70,148                     & 69,066   \\
Frame time (s)       & 3.1                        & 3.2 \\
\enddata
\tablenotetext{a}{Nominal pointing position was 
(J2000) R.A. = 20$h$ 21$m$ 46.42$s$, 
decl. = $+$37$^{\circ}$ 22$'$ 44.9$''$.}
\tablenotetext{b}{Nominal pointing position was 
(J2000) R.A. = 20$h$ 21$m$ 38.39$s$,
decl. = $+$37$^{\circ}$ 30$'$ 04.1$''$.}
\tablenotetext{c}{ACIS-I consists of four 1024 $\times$ 1024 pixel
CCDs with a pixel size of 0$''$.492. The combined field-of-view is
16.9$'$ $\times$ 16.9$'$.
The energy range is E $\approx$ 0.5 - 10 keV.
For an on-axis  point source the 90\% encircled energy radius is 
R$_{90}$ $\approx$ 0$''$.9.
The energy resolution at 1.49 keV is $\approx$0.13 keV.
}
\end{deluxetable}

\clearpage

%% file: table2.tex
\begin{deluxetable}{lllcccccl}
\tabletypesize{\scriptsize}
\tablewidth{0pt} 
\tablecaption{Chandra X-Ray Sources Near G75.77$+$0.34}
\tablehead{
           \colhead{Nr.}               &
           \colhead{R.A.}               &
           \colhead{Decl.}              &
           \colhead{Net Counts}         &
           \colhead{E$_{50}$}           &
           \colhead{Hardness}           &
           \colhead{P$_{var}$}          &
           \colhead{K}        &
           \colhead{Identification (offset)}     \\    
           \colhead{}                   & 
           \colhead{(J2000)}            &
           \colhead{(J2000)}            &
           \colhead{(cts)}              &                            
           \colhead{(keV)}              &
           \colhead{}                   &         
           \colhead{}                   &
           \colhead{(mag)}              & 
           \colhead{UGPS ... (arcsec)} 
                                  }
\startdata
1  & 20 21 38.72  & +37 26 02.2 &  11$\pm$4 & 2.58 & 0.69  & 0.36 &  13.99     &  J202138.73+372602.1 (0.23)\tablenotemark{b}   \\
2  & 20 21 39.24  & +37 25 33.5 &   8$\pm$3 & 3.27 & 0.82  & 0.47 &  14.30     &  J202139.21+372533.9 (0.51)\tablenotemark{b}   \\
3  & 20 21 39.47  & +37 25 53.6 &  31$\pm$6 & 1.81 & 0.29  & 0.37 &  13.98     &  J202139.46+372553.6 (0.07)\tablenotemark{b}   \\
4  & 20 21 40.09  & +37 26 04.9 &  25$\pm$5 & 2.09 & 0.53  & 0.27 &  14.16     &  J202140.09+372604.9 (0.04) \\
5  & 20 21 40.43  & +37 26 04.9 &  43$\pm$7 & 2.85 & 0.82  & 0.25 &  13.94     &  J202140.43+372604.8 (0.10) \\
6  & 20 21 40.54  & +37 25 48.3 &  54$\pm$8 & 3.11 & 0.90  & 0.41 &  13.37     &  J202140.54+372548.4 (0.15)\tablenotemark{b,c} \\
7  & 20 21 40.74  & +37 26 19.6 &  34$\pm$6 & 3.06 & 0.72  & 0.99 &  15.23     &  J202140.75+372619.9 (0.36)     \\
8  & 20 21 40.75  & +37 25 33.4 &  61$\pm$8 & 3.60 & 0.92  & 0.99 &  12.88     &  J202140.69+372533.1 (0.70) \\
9  & 20 21 40.97  & +37 26 16.5 &  38$\pm$7 & 2.83 & 0.69  & 0.31 &  12.97     &  J202140.99+372616.5 (0.30)\tablenotemark{b} \\
10 & 20 21 41.02  & +37 25 50.2 &  20$\pm$5 & 3.36 & 0.74  & 0.59 &  13.88     &  J202141.01+372550.0 (0.15)\tablenotemark{b,d} \\
11 & 20 21 41.18  & +37 25 59.7 &  13$\pm$4 & 2.90 & 0.89  & 0.72 &  13.78     &  J202141.17+372559.6 (0.09) \\
12 & 20 21 41.71  & +37 25 58.9 &  26$\pm$6 & 3.37 & 0.88  & 0.70 &  13.47     &  J202141.70+372559.0 (0.10)\tablenotemark{b} \\
13 & 20 21 41.83  & +37 25 50.0 &  55$\pm$8 & 3.49 & 0.90  & 0.99 &  13.58     &  J202141.81+372550.3 (0.41)\tablenotemark{b} \\
14 & 20 21 42.18  & +37 25 48.4 &  10$\pm$4 & 1.72 & 0.38  & 0.48 &  13.96     &  J202142.19+372548.7 (0.43)\tablenotemark{b} \\
15 & 20 21 42.76  & +37 25 48.4 &  13$\pm$4 & 3.37 & 0.71  & 0.24 &  14.84     &  J202142.80+372548.3 (0.51) \\
\enddata
\tablenotetext{a}{
Notes:~Tabulated sources lie within a 1$'$ $\times$ 1$'$ region centered on G75.77$+$0.34 (Fig. 3).
X-ray data are from ObsId 9914 (except CXO J202140.74$+$372619.6 which was only detected in
ObsId 18083) using events in the 0.2 - 8 keV range lying  within the position 
error ellipse (3$\sigma$) of each source. Only sources with $\geq$5 net counts are included.
Tabulated quantities are: J2000.0 X-ray position (R.A., decl.); net counts and 
net counts error from {\em wavdetect} (rounded to the nearest integer,
background subtracted and PSF-corrected);  
median photon energy E$_{50}$; Hardness =  counts(2-8 keV)/counts(0.2-8 keV);
probability that the source is variable P$_{var}$ as determined from
CIAO $glvary$; K magnitude of UGPS counterpart (AperMag3 value); 
UGPS counterpart name within a 1$''$ search radius centered on the X-ray position.
Typical K-band photometry errors given in the UKIDSS UGPS archives for
the tabulated sources are $\pm$0.005 [range 0.002 - 0.014] mag.
The offset (in parentheses) is given in arcseconds between the X-ray and UGPS position. 
}
\tablenotetext{b}{A 2MASS counterpart is also found within 1$''$ of the X-ray position.}
\tablenotetext{c}{Possible counterpart of radio peak G75.77$+$0.34B.}
\tablenotetext{d}{A second possible counterpart UGPS J202140.98$+$373550.4 
                  (offset = 0$''$.54; H = 15.18) is also found but is not detected at 
                  J or K bands.}
\end{deluxetable}

%% file: table3.tex
\begin{deluxetable}{lllcccccl}
\tabletypesize{\scriptsize}
\tablewidth{0pt} 
\tablecaption{Chandra X-Ray Sources Near G75.84$+$0.40}
\tablehead{
           \colhead{Nr.}               &
           \colhead{R.A.}               &
           \colhead{Decl.}              &
           \colhead{Net Counts}         &
           \colhead{E$_{50}$}           &
           \colhead{Hardness}           &
           \colhead{P$_{var}$}          &
           \colhead{K$_{s}$}            &
           \colhead{Identification (offset)\tablenotemark{b}}     \\    
           \colhead{}                   &
           \colhead{(J2000)}            &
           \colhead{(J2000)}            &
           \colhead{(cts)}              &                            
           \colhead{(keV)}              &
           \colhead{}                   &         
           \colhead{}                   &
           \colhead{(mag)}              & 
           \colhead{(arcsec)} 
                                  }
\startdata
1  & 20 21 36.28  & +37 31 24.0 &  7$\pm$3 & 2.16  & 0.57 & 0.38  &            &       \\
2  & 20 21 37.31  & +37 31 13.0 & 12$\pm$4 & 2.87  & 0.77 & 0.55  &            &         \\
3  & 20 21 37.36  & +37 31 37.7 & 47$\pm$7 & 3.06  & 0.79 & 0.12  &            &         \\
4  & 20 21 37.69  & +37 31 15.1 & 19$\pm$5 & 3.40  & 0.82 & 0.59  &            & G75.84+0.40B radio (3.8) \\
5  & 20 21 37.95  & +37 31 18.6 & 10$\pm$3 & 3.67  & 0.75 & 0.81  &            & \\
6  & 20 21 38.01  & +37 31 14.5 & 50$\pm$7 & 3.15  & 0.72 & 0.35  &   9.63     & 2M J202137.97+373115.2 (0.84)\tablenotemark{c} \\
7  & 20 21 38.30  & +37 31 24.2 & 84$\pm$9 & 3.47  & 0.90 & 1.00  &            & \\
8  & 20 21 38.38  & +37 31 15.8 & 24$\pm$5 & 3.78  & 0.85 & 0.82  &            & \\
9  & 20 21 38.98  & +37 31 14.0 & 47$\pm$7 & 3.60  & 0.86 & 0.17  &            & \\
10 & 20 21 39.35  & +37 31 18.8 &  8$\pm$3 & 2.47  & 0.57 & 0.51  &   12.59    & UGPS J202139.23+373119.2 (1.38)\tablenotemark{d} \\ 
11 & 20 21 40.00  & +37 31 17.1 &  8$\pm$3 & 3.00  & 0.78 & 0.65  &            & \\
12 & 20 21 40.32  & +37 31 32.2 &  9$\pm$3 & 3.79  & 1.00 & 0.46  &            & \\ 
13 & 20 21 40.71  & +37 31 40.7 &  8$\pm$3 & 1.69  & 0.33 & 0.36  &  13.31     & UGPS J202140.70+373140.9 (0.24)\tablenotemark{e}  \\
14 & 20 21 40.95  & +37 30 53.7 &  7$\pm$3 & 1.69  & 0.29 & 0.36  &  13.60     & UGPS J202140.92+373053.9 (0.36)\tablenotemark{f}  \\ 
\enddata
\tablenotetext{a}{
Notes -- Tabulated sources lie within a 1$'$ $\times$ 1$'$ region centered on G75.84$+$0.40 (Fig. 5).
X-ray data are from ObsId 18083 using events in the 0.2 - 8 keV range lying  within the 
position error ellipse (3$\sigma$) of each source. Only sources with $\geq$5 net counts are included.
Tabulated quantities are:  J2000.0 X-ray position (R.A., decl.); net counts and 
net counts error from {\em wavdetect} (rounded to the nearest integer,
background subtracted and PSF-corrected);  
median photon energy E$_{50}$; Hardness = counts(2-8 keV)/counts(0.2-8 keV);
probability that the source is variable P$_{var}$ as determined from CIAO $glvary$;
2MASS K$_{s}$ or UGPS K (AperMag3) magnitude of the near-IR counterpart;
and 2MASS (2M), UGPS, or {\em HST} Guide Star Catalog (GSC v2.3) 
counterpart name within a 1$''$ search radius centered on the X-ray position.
The offset (in parentheses) is given in arcseconds between the X-ray and counterpart position. 
}

\tablenotetext{b}{Identification of IR counterparts in this region and their photometric quality
                  is adversely affected by high background
                  emission from the M-type supergiant BC Cyg lying 48$''$ to the north.
                  Typical K-band photometry errors given in the UKIDSS UGPS archives for 
                  the tabulated sources  are $\pm$0.004 [range 0.004 - 0.005] mag.} 
\tablenotetext{c}{Detected by 2MASS only at K$_{s}$ band (K$_{s}$ = 9.628$\pm$0.029)
                  and confusion flag is set. No UGPS counterpart found.}
\tablenotetext{d}{Detected only at K band. Offset = 1$''$.38 so match is questionable.}
\tablenotetext{e}{{\em HST} GSC202140.72+373141.1 (0.48), class=3 (non-stellar).}
\tablenotetext{f}{{\em HST} GSC202140.93+373053.7 (0.24), class=3 (non-stellar).} 
\end{deluxetable}

%% file: table4.tex
\begin{deluxetable}{lccc}
\tabletypesize{\footnotesize}
\tablecaption{Spectral Fits for X-ray Sources Near H II Radio Peaks}
\tablehead{
\colhead{Parameter}        &
\colhead{G75.77$+$0.34B}   &
\colhead{G75.84$+$0.40B}   &
\colhead{G75.84$+$0.40Bb}   
}
\startdata
CXO source id.                                   & J202140.54$+$372548.3 & J202137.69+373115.1   & J202138.01+373114.5    \\
ObsId                                            & 9914                  & 18083                 & 18083      \\
Model                                            & 1T APEC               & 1T APEC               & 1T APEC  \\
Abundances                                       & solar                 & solar                 & solar   \\
N$_{\rm H}$ (10$^{22}$ cm$^{-2}$)                & 5.06 [3.65 - 7.07]    & 1.84 [0.65 - 3.46]    & 1.63 [1.01 - 2.71]  \\
kT (keV)                                         & 1.91 [1.32 - 3.05]    & 6.88 [2.60 - ...]     & 13.3 [3.07 - ...]  \\
norm (10$^{-5}$ cm$^{-5}$)                       & 5.55 [2.99 - 11.7]    & 0.59 [0.35 - 1.12]    & 1.00 [0.78 - 1.41] \\
$\chi^2$/dof                                     & 2.25/8                & 2.04/3                & 4.62/7     \\
$\chi^2_{\rm red}$                               & 0.28                  & 0.68                  & 0.66 \\
Net counts                                       & 54$\pm$8              & 19$\pm$5              & 50$\pm$7  \\
Hardness ratio                                   & 0.90                  & 0.82                  & 0.72 \\
Counts/bin                                       & 5                     & 4                     & 4  \\
F$_{\rm X}$ (10$^{-14}$ ergs cm$^{-2}$ s$^{-1}$) & 1.47 (9.06)           & 0.62 (1.12)           & 1.14 (1.87) \\
log L$_{\rm X}$ @ d = 3.5 kpc (ergs s$^{-1}$)    & 32.12                 & 31.21                 &  31.44  \\
\enddata
\tablecomments{
Based on  fits of binned ACIS-I spectra  using XSPEC v12.8.2.
Square brackets enclose 1$\sigma$ confidence intervals. 
An ellipsis means the algorithm used to compute confidence limits did not converge.
Net counts are from CIAO  {\em wavdetect}.
Hardness ratio = counts(2-8 keV)/counts(0.2-8 keV).
The X-ray flux (0.3-8 keV) is the absorbed value followed in parentheses by the 
unabsorbed value. For $apec$ models, the volume emission measure 
is related to the normalization (norm) by
n$_{e}^2$V = 4$\pi$$\times$10$^{14}$d$_{cm}^2$$\cdot$norm,
where n$_{e}$ is electron density, V is the volume of 
X-ray emitting plasma, and d$_{cm}$ is the distance in cm.
The X-ray luminosity (L$_{x}$) is computed 
at an assumed distance of 3.5 kpc (Xu et al. 2013).
Solar abundances are from Anders \& Grevesse (1989).
}
\end{deluxetable}

%% file: table5.tex
\begin{deluxetable}{llll}
\tabletypesize{\scriptsize}
\tablewidth{0pc}
\tablecaption{{\em Chandra} X-ray Spectral Fits for WR 142
\label{tbl-1}}
\tablehead{
\colhead{Parameter}      &
\colhead{       }        &
\colhead{       }        &
\colhead{       }
}
\startdata
Model                                                    & 1T APEC               & 1T APEC                       & POW$+$GAUSS       \\
N$_{\rm H}$ (10$^{22}$ cm$^{-2}$)                        & 10.7 [7.0-19.5]       & 10.8 [6.9-16.3]               & 5.3 [0.8-16.5]    \\
kT (keV)                                                 & 5.5 [2.2-24.5]        & \{5.5\}\tablenotemark{a}      & ...               \\
norm (cm$^{-5}$)                                         & 3.1 [1.9-11.2]e-05    & 7.3 [5.0-10.6]e-13            & 1.5 [0.4-4.4]e-06 \\
$\Gamma_{\rm ph}$\tablenotemark{b}                       & ...                   & ...                           & 0.76 [$-$0.06-$+$1.5]  \\
E$_{\rm line}$ (keV)                                     & ...                   & ...                           & 6.65 [6.06-6.91]\tablenotemark{c} \\
norm$_{\rm line}$ (cm$^{-5}$)                            & ...                   & ...                           & 4.0 [0.6-7.3]e-07 \\
Abundances\tablenotemark{d}                              & solar                 & WO                            & ...               \\
$\chi^{2}$/dof                                           & 19.25/26              & 20.67/27                      & 18.43/25          \\
reduced $\chi^{2}$                                       & 0.74                  & 0.77                          & 0.74              \\
F$_{\rm x}$ (10$^{-14}$ ergs cm$^{-2}$ s$^{-1}$)\tablenotemark{e}  & 1.65 (5.71)           & 1.42 (5.39)                   & 1.98 (2.97)       \\
F$_{\rm x,line}$ (10$^{-14}$ ergs cm$^{-2}$ s$^{-1}$)    & ...                   & ...                           & 0.39 (0.43)       \\
log L$_{\rm x}$ (ergs s$^{-1}$)                          & 31.32                 & 31.29                         & 31.03             \\
log (L$_{\rm x}$/L$_{bol}$)\tablenotemark{f}             & $-$7.76               & $-$7.79                       & $-$8.05           \\
\enddata
\tablecomments{
Based on  simultaneous fits of background-subtracted ACIS-I spectra from 
ObsId 9914 (46 net counts) and ObsId 18083 (26 net counts) binned to a 
minimum of 5 counts per bin using XSPEC v12.8.2. Models include an absorption
component (N$_{\rm H}$) and are single-temperature thermal plasma (1T APEC) 
and  power-law plus Gaussian line (POW$+$GAUSS). The line is added to
reproduce the Fe K feature near 6.7 keV. The tabulated parameters
are absorption column density (N$_{\rm H}$), plasma temperature (kT),
XSPEC normalization (norm), photon power-law index ($\Gamma_{\rm ph}$), 
For $apec$ models, the volume emission measure 
is related to the normalization (norm) by
n$_{e}^2$V = 4$\pi$$\times$10$^{14}$d$_{cm}^2$$\cdot$norm,
where n$_{e}$ is electron density, V is the volume of 
X-ray emitting plasma, and d$_{cm}$ is the distance in cm.
Gaussian line centroid energy (E$_{\rm line}$), line normalization (norm$_{\rm line}$).
Quantities enclosed in curly braces were held fixed during fitting.
Square brackets enclose 1$\sigma$ confidence intervals.
X-ray flux (F$_{\rm X}$) in the 0.3 - 8 keV range is the  observed (absorbed) 
value followed in parentheses by the unabsorbed value.
The continuum-subtracted Gaussian line flux (F$_{\rm X,line}$)
is measured in the 6.4 - 6.9 keV range.
X-ray luminosity (L$_{\rm X}$) is the unabsorbed value in the
0.3 - 8.0 keV range. A distance of 1.737 kpc is assumed ({\em Gaia} DR2).}
\tablenotetext{a}{Temperature held fixed at the best-fit value of the solar abundance model.
If kT is allowed to vary using WO abundances it runs away to the maximum temperature of 64 keV 
allowed by the APEC model.}
\tablenotetext{b}{A photon power-law index $\Gamma_{\rm ph}$ corresponds to flux scaling
with photon energy E according to 
F$_{\rm x}$ $\propto$ E$^{-(\Gamma_{\rm ph} -1)}$.}
\tablenotetext{c}{~Line width held fixed at 0.12 keV (FWHM).}
\tablenotetext{d}{Solar abundances are from Anders \& Grevesse (1989).
WO abundances are from Table 1 of van der Hucht et al. (1986),
except the hydrogen and nitrogen abundances are arbitrarily set to the small
non-zero value 1 $\times$ 10$^{-6}$ for compatibility with XSPEC.}
\tablenotetext{e}{Measurement uncertainties in observed (absorbed) flux for the simultaneous
                  fits of both spectra are $\approx$20\% (1$\sigma$).}
\tablenotetext{f}{Assumes log (L$_{bol}$/L$_{\odot}$) = 5.50$\pm$0.15 (Oskinova et al. 2009; 
                  Tramper et al. 2015).}

\end{deluxetable}

%% file: table6.tex
\begin{deluxetable}{lccc}
\tabletypesize{\footnotesize}
\tablecaption{Spectral Fits for Diffuse Emission Near H II Regions}
\tablehead{
\colhead{Parameter}        &
\colhead{G75.77$+$0.34}   &
\colhead{G75.84$+$0.40}   &
\colhead{G75.84$+$0.40}   
}
\startdata
ObsId                                            & 9914                  & 18083                 & 18083      \\
Region (arcsec$^2$)                              & ellipse (867)         & circle (616)          & circle (616) \\
Model                                            & 1T APEC               & 1T APEC               & POWER LAW  \\
Abundances                                       & solar                 & solar                 & ...    \\
N$_{\rm H}$ (10$^{22}$ cm$^{-2}$)                & 3.6 [2.8 - 4.9]       & 4.2 [3.2 - 6.3]       & 6.2 [4.3 - 8.5]  \\
kT (keV)                                         & 5.4 [3.3 - 9.1]       & 2.2 [1.5 - 2.8]       & ...  \\
$\Gamma_{ph}$                                    & ...                   & ...                   & 3.9 [3.0 - 4.9] \\
norm (10$^{-4}$ cm$^{-5}$)                       & 5.55 [2.99 - 11.7]    & 1.14                  & 3.99  \\
$\chi^2$/dof                                     & 20.19/20              & 27.43/30              & 22.98/30     \\
$\chi^2_{\rm red}$                               & 1.01                  & 0.91                  & 0.77 \\
Net counts                                       & 160                   & 131                   & 131  \\
F$_{\rm X},diffuse$ (10$^{-14}$ ergs cm$^{-2}$ s$^{-1}$) & 4.99 (11.4)           & 3.98 (11.8)           & 3.67 (323.0) \\
log L$_{\rm X},diffuse$ (ergs s$^{-1}$)                  & 32.22                 & 32.42                 & 33.67  \\
\enddata
\tablecomments{
Based on  fits of binned ACIS-I spectra  using XSPEC v12.8.2.
Spectral extraction regions and spectra are shown in Figure 9.
Square brackets enclose 1$\sigma$ confidence intervals. 
The X-ray flux (0.3 - 8 keV) is the absorbed value followed in parentheses by the 
unabsorbed value. For $apec$ models, the volume emission measure 
is related to the normalization (norm) by
n$_{e}^2$V = 4$\pi$$\times$10$^{14}$d$_{cm}^2$$\cdot$norm,
where n$_{e}$ is electron density, V is the volume of 
X-ray emitting plasma, and d$_{cm}$ is the distance in cm.
A photon power-law index $\Gamma_{\rm ph}$ corresponds to flux scaling
with photon energy E according to
F$_{\rm x}$ $\propto$ E$^{-(\Gamma_{\rm ph} -1)}$.
A distance of 3.5 kpc is assumed. Solar abundances are from Anders \& Grevesse (1989).
}
\end{deluxetable}

%% file: table7.tex
\begin{deluxetable}{lccclll}
\tabletypesize{\scriptsize}
\tablewidth{0pt} 
\tablecaption{Selected UGPS Counterparts to Chandra Detections in ON 2\tablenotemark{a}}
\tablehead{
           \colhead{UGPS name}  &
           \colhead{K}          &
           \colhead{H-K}        &
           \colhead{J-H}        &
           \colhead{Chandra position}   &
           \colhead{Net cts}  &
           \colhead{E$_{50}$} \\
           \colhead{}    &
           \colhead{(mag)}    &
           \colhead{(mag)}    &
           \colhead{(mag)}    &
           \colhead{R.A., decl.} &
           \colhead{(cts)}       &
           \colhead{(keV)}
}
\startdata
 J202116.32+373339.4                  &14.470  &0.337  &0.516  &J20 21 16.38 +37 33 38.9 & 10$\pm$4   & 1.21  \\
 J202117.36+372437.3                  &18.084  &1.283  &0.446  &J20 21 17.37 +37 24 36.7 & 38$\pm$7   & 2.05 \\
 J202119.77+372935.7\tablenotemark{b} &16.193  &0.883  &1.067  &J20 21 19.79 +37 29 35.3 & 6$\pm$3    & 2.02 \\
 J202124.19+372816.8                  &12.340  &0.402  &0.604  &J20 21 24.20 +37 28 16.6 & 10$\pm$3   & 2.26 \\
 J202128.16+373316.4\tablenotemark{b} &11.737  &0.544  &0.863  &J20 21 28.19 +37 33 16.5 & 43$\pm$7   & 3.19 \\
 J202128.23+373158.5\tablenotemark{b} &14.607  &0.728  &0.969  &J20 21 28.27 +37 31 58.0 & 142$\pm$12 & 2.48\tablenotemark{c}  \\
 J202133.20+372951.3\tablenotemark{b} &15.110  &0.975  &1.538  &J20 21 33.20 +37 29 51.0 & 8$\pm$3    & 3.04 \\
 J202134.67+372344.8\tablenotemark{b} &13.324  &1.502  &2.236  &J20 21 34.67 +37 23 44.9 & 20$\pm$5   & 3.73 \\
 J202134.90+372442.0\tablenotemark{b} &13.524  &2.620  &3.697  &J20 21 34.91 +37 24 42.3 & 29$\pm$6   & 3.65\tablenotemark{c} \\
 J202139.56+372634.1\tablenotemark{b} &12.673  &1.504  &2.219  &J20 21 39.61 +37 26 34.1 & 51$\pm$7   & 3.34 \\
 J202139.75+372818.0                  &12.726  &0.318  &0.537  &J20 21 39.77 +37 28 17.5 & 9$\pm$3    & 2.20 \\
 J202141.01+372550.0                  &13.875  &1.593  &1.049  &J20 21 41.02 +37 25 50.2 & 20$\pm$5   & 3.36 \\
 J202141.29+373401.0\tablenotemark{b} &16.574  &1.019  &1.544  &J20 21 41.30 +37 34 00.7 & 14$\pm$4   & 2.16 \\
 J202142.66+373031.4\tablenotemark{b} &14.525  &1.768  &2.969  &J20 21 42.70 +37 30 31.3 & 17$\pm$4   & 3.86\tablenotemark{c} \\
 J202144.08+372628.1\tablenotemark{b} &16.744  &1.093  &1.491  &J20 21 44.14 +37 26 28.4 & 22$\pm$5   & 3.24\tablenotemark{c} \\
 J202146.41+372653.4\tablenotemark{b} &15.524  &0.856  &1.304  &J20 21 46.41 +37 26 54.2 & 4$\pm$2\tablenotemark{d} & 4.12 \\
 J202147.57+373418.9                  &11.828  &0.231  &0.321  &J20 21 47.14 +37 34 18.2 & 51$\pm$7   & 2.01 \\           
 J202150.33+373005.0\tablenotemark{b} &15.631  &0.801  &1.007  &J20 21 50.33 +37 30 04.6 & 5$\pm$2    & 1.74 \\
 J202150.55+373327.6\tablenotemark{b} &13.317  &0.506  &0.839  &J20 21 50.54 +37 33 27.1 & 14$\pm$4   & 2.82 \\
 J202151.43+372916.0                  &11.792  &0.338  &0.465  &J20 21 51.39 +37 29 15.6 & 10$\pm$3   & 2.16 \\
\enddata
\tablenotetext{a}{UGPS name identifies the UKIDSS GPS DR6 (UGPS) near-IR source with detections in all three
                  bands lying within 1$''$ of the  Chandra source position. Only UGPS sources
                  with colors lying to the right (redward) of the A0V line in Figure 10 are listed.
                  Magnitudes and colors are based on UGPS DR6 AperMag3 values.
                  Typical photometry errors given in the UKIDSS UGPS archives for the tabulated
                  sources and their ranges are J: $\pm$0.016 [0.001 - 0.194], H: $\pm$0.009 [0.001 - 0.264],
                  K: $\pm$0.008 [0.001 - 0.172] mag. Net counts are measured in the 0.2-8 keV range.
                  E$_{50}$ is median X-ray photon energy.}
\tablenotetext{b}{Near-IR colors are consistent with classical T Tauri stars.}
\tablenotetext{c}{CIAO {\em glvary} returns a high X-ray source variability probability: P$_{var}$ $\geq$ 0.97.}
\tablenotetext{d}{Low significance detection.}
\end{deluxetable}

%% file: f1.tex
\begin{figure}
\figurenum{1}
\includegraphics*[width=12.0cm,angle=0]{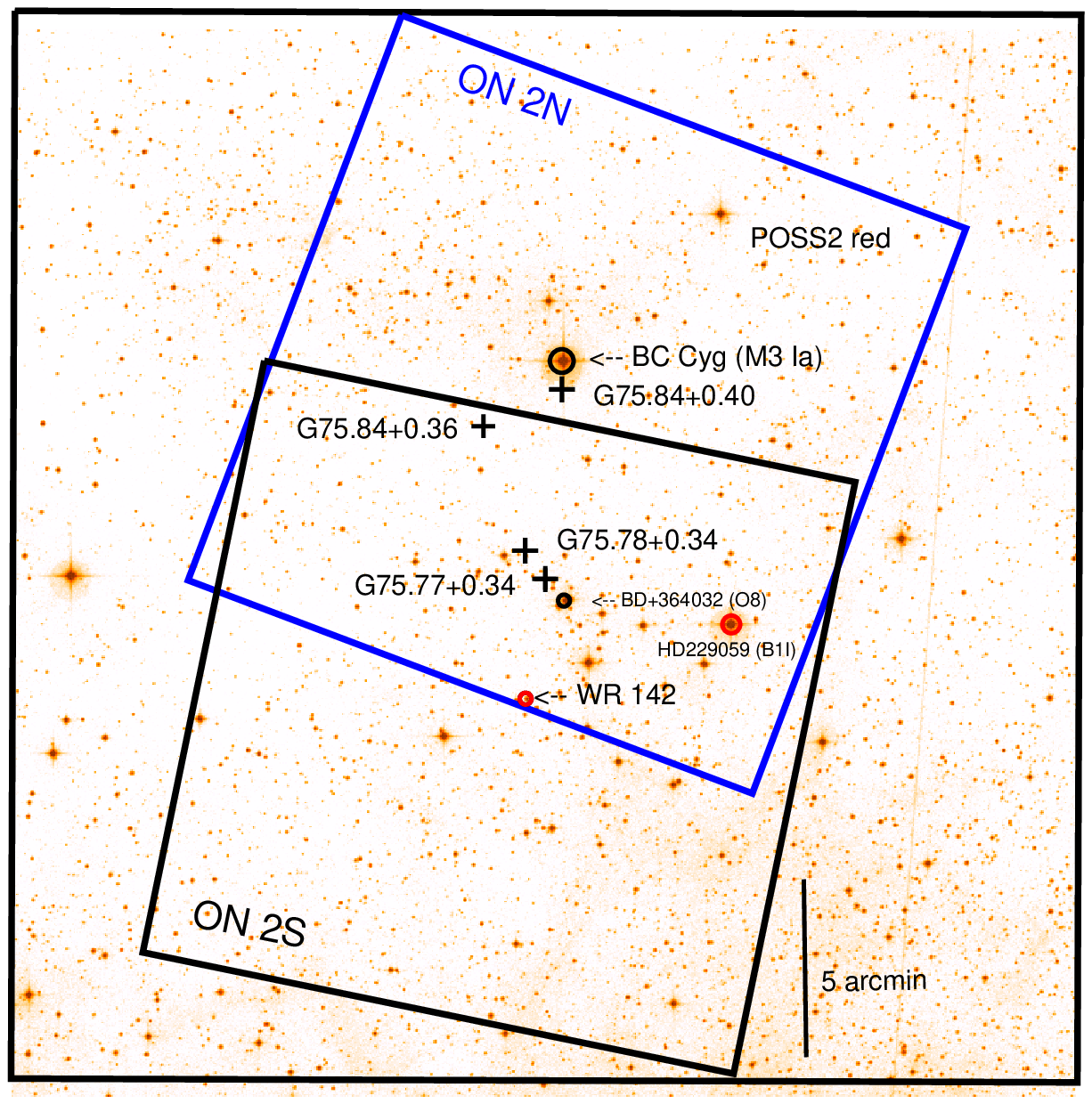}
\caption{POSS2 red image of ON 2 with {\em Chandra} ACIS-I
detector footprints (17$'$ $\times$ 17$'$) overlaid. 
Radio positions of HII regions are shown. 
Positions of three optically-identified members of
the Berkeley 87 cluster are marked (BD$+$36 4032,
HD 229059, WR 142). 
The M3 supergiant BC Cyg is IR-bright but is not  
detected in X-rays. The outer bounding box has
dimensions 30$'$ $\times$ 30$'$. N is up, E is left.
}
\end{figure}

%% file: f2.tex
\begin{figure}
\figurenum{2}
\includegraphics*[width=8.0cm,angle=0]{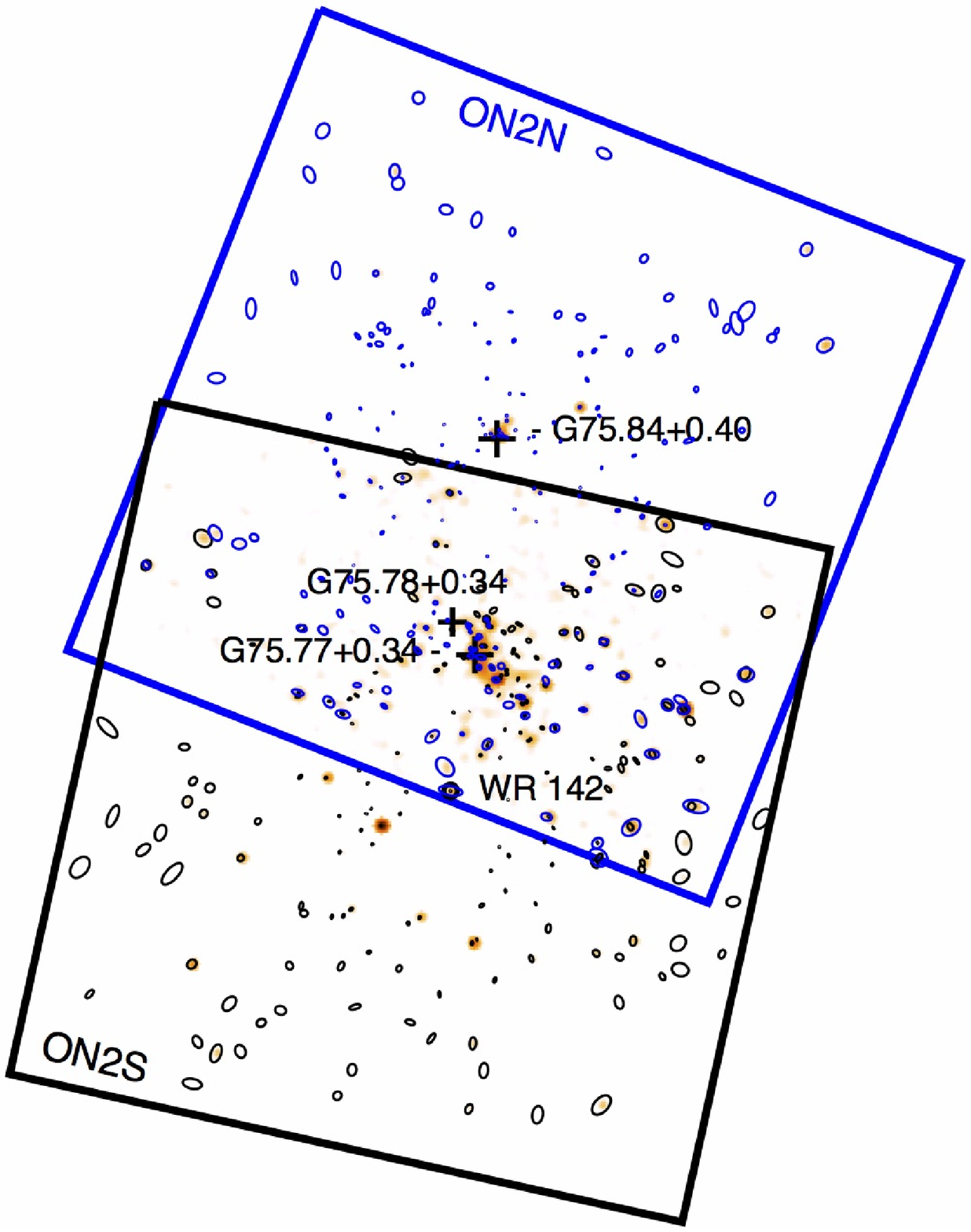}
\includegraphics*[width=8.0cm,angle=0]{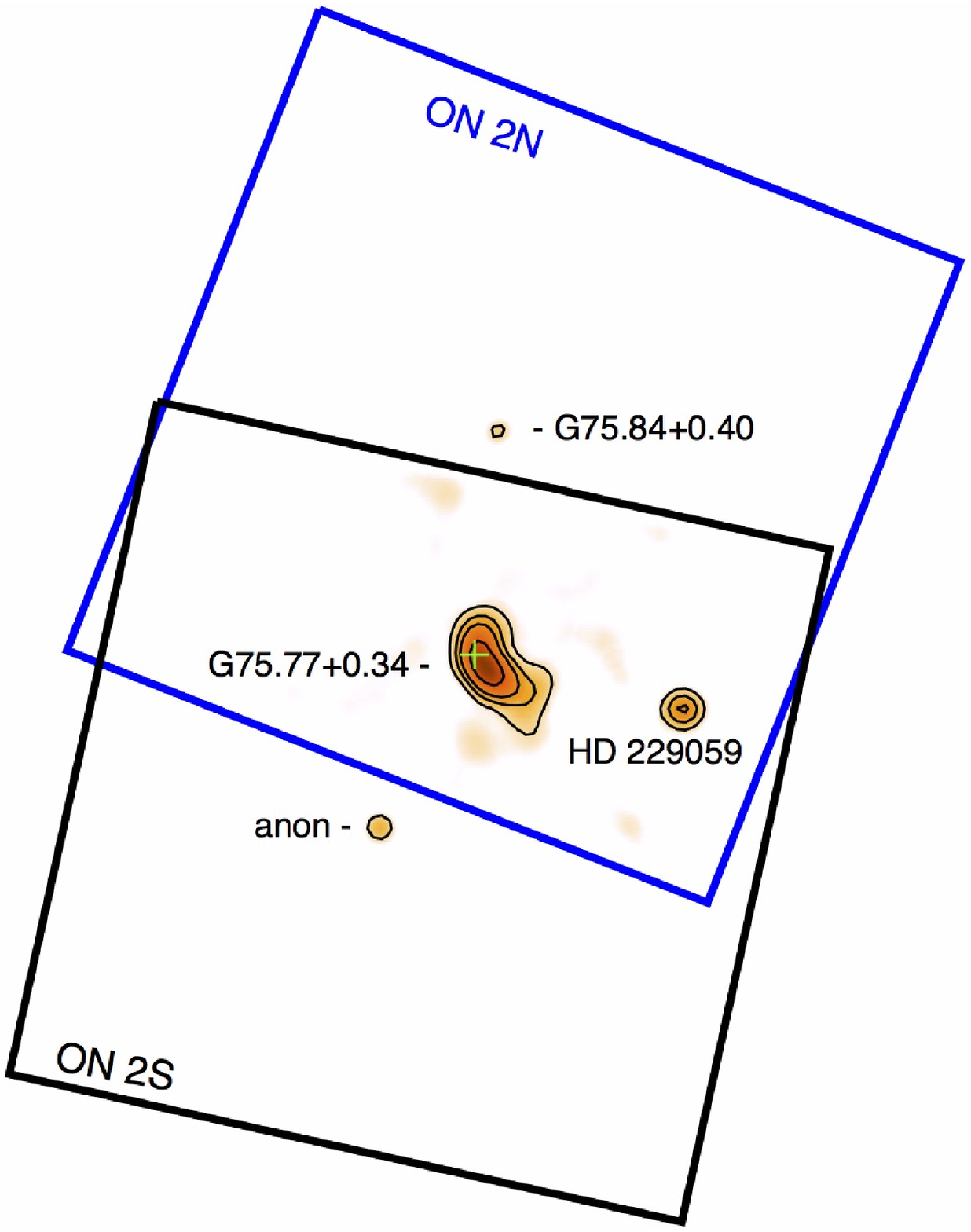}
\caption{{\em Left}:~Mosaiced {\em Chandra} ACIS-I images (0.2 - 8 keV) of ON2 from ObsId 9914 (ON2S) 
and ObsId 18083 (ON2N). Ellipses (3$\sigma$) mark  {\em Chandra} X-ray source positions. 
Crosses ($+$) show HII region mean radio continuum positions. WR 142 lies in the 
overlap region and was detected in both observations. N is up and E is left.
{\em Right}:~Same as at left but heavily binned and smoothed to bring out 
the regions with brightest X-ray emission. X-ray properties of the B1 Ia
star HD 229059 were given in Sokal et al. (2010).
No optical or 2MASS counterpart was found for the X-ray source marked `anon'
(CXO J202152.99$+$372139.33) but it is coincident with a faint
{\em Spitzer} mid-IR source (Sec. 3.7). The image is binned by a factor of 8
(binned pixel size = 3.$''$94). Contours are at (4,5,6,8,10) counts
per binned pixel. 
}
\end{figure}

%% file: f3.tex
\begin{figure}
\figurenum{3}
\includegraphics*[width=8.0cm,angle=0]{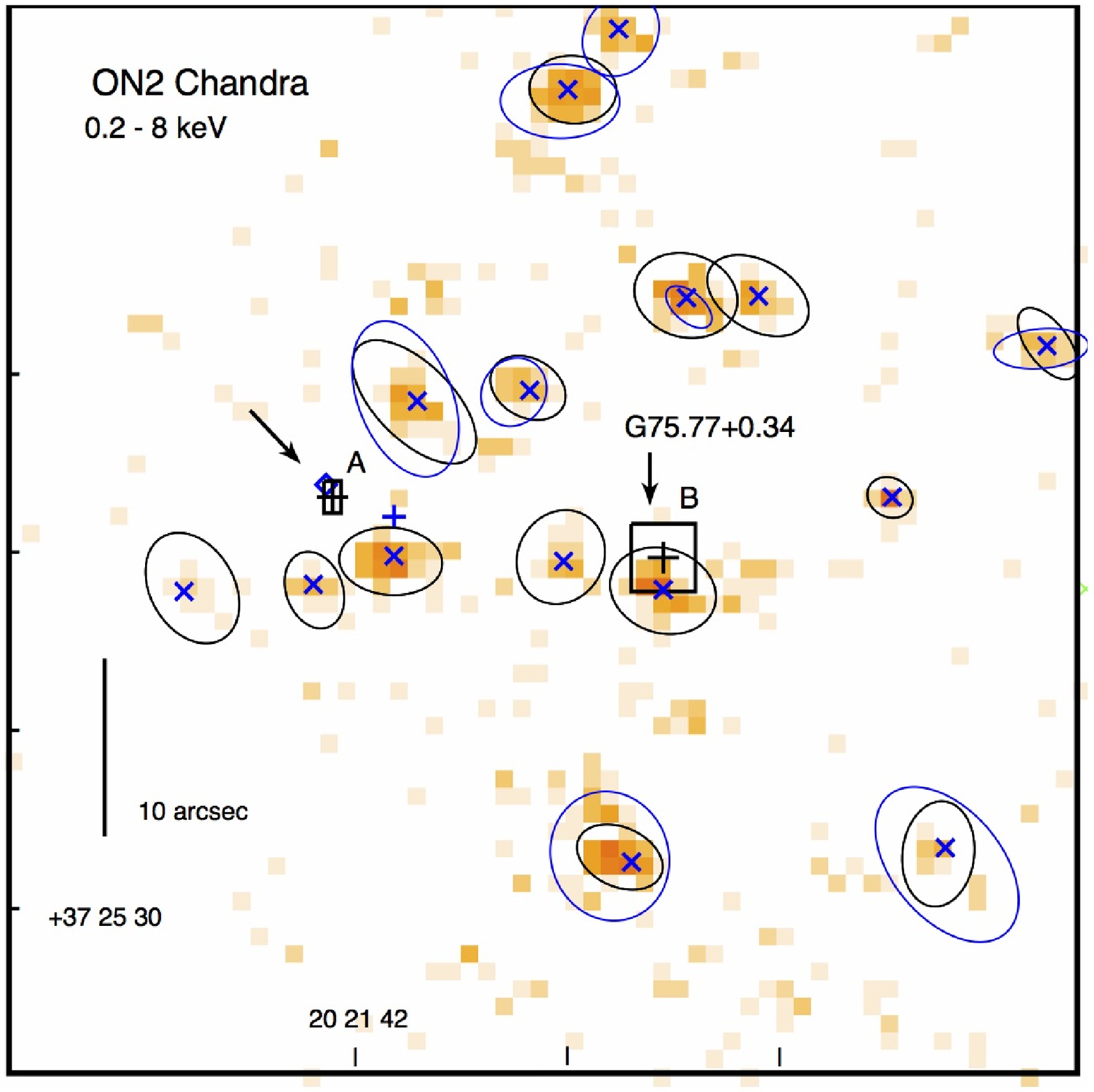} \\
\includegraphics*[height=7.5cm,angle=-90]{f3b.eps}
\caption{{\em Top}:~Mosaiced {\em Chandra} ACIS-I image (0.2 - 8 keV) 
of a 1$'$ $\times$ 1$'$ region  near G75.77$+$0.34, combining data from
both observations. {\em Chandra} source position error
ellipses (3$\sigma$) are shown in black (ObsId 9914) and blue (ObsId 18083).
The $\times$ symbols mark UGPS near-IR counterpart positions (Table 2).
Rectangles are position error boxes 
for Westerbork radio continuum  sources G75.77$+$0.34A and B from Matthews et al. (1973).
The diamond overlapping peak A marks the  position of 2MASS J202142.13$+$372554.3.
The  position of {\em Chandra} source J202140.54$+$372548.3 overlaps that of
Westerbork radio source B but there is no X-ray detection of radio source A.
The blue cross ($+$) southwest of  G75.77$+$0.34A marks the VLA radio peak position of 
Garay et al. (1993), lying just outside the position error ellipse of CXO J202141.83$+$372550.0.
Log intensity scale. Coordinate tick marks are J2000. 
{\em Bottom}:~{\em Chandra} ACIS-I spectrum of the X-ray source coincident
with radio peak G75.77$+$0.34B (CXO J202140.54$+$372548.3), 
binned to a minimum of 5 counts per bin.
The dashed line is a 1T  thermal plasma model fit (Table 4).
}
\end{figure}

%% file: f4.tex
\begin{figure}
\figurenum{4}
\includegraphics*[width=9.0cm,angle=0]{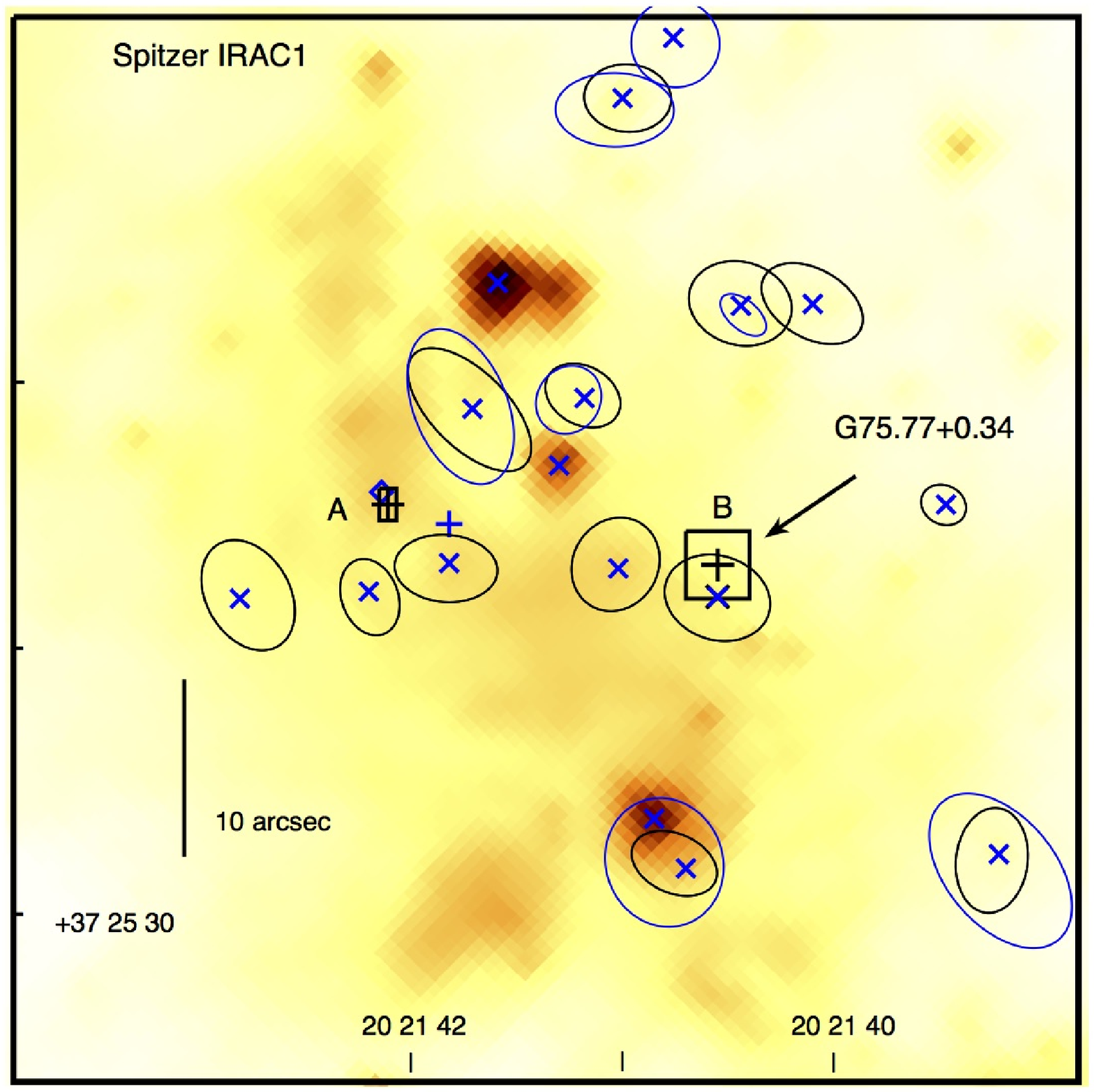} \\
\includegraphics*[width=9.0cm,angle=0]{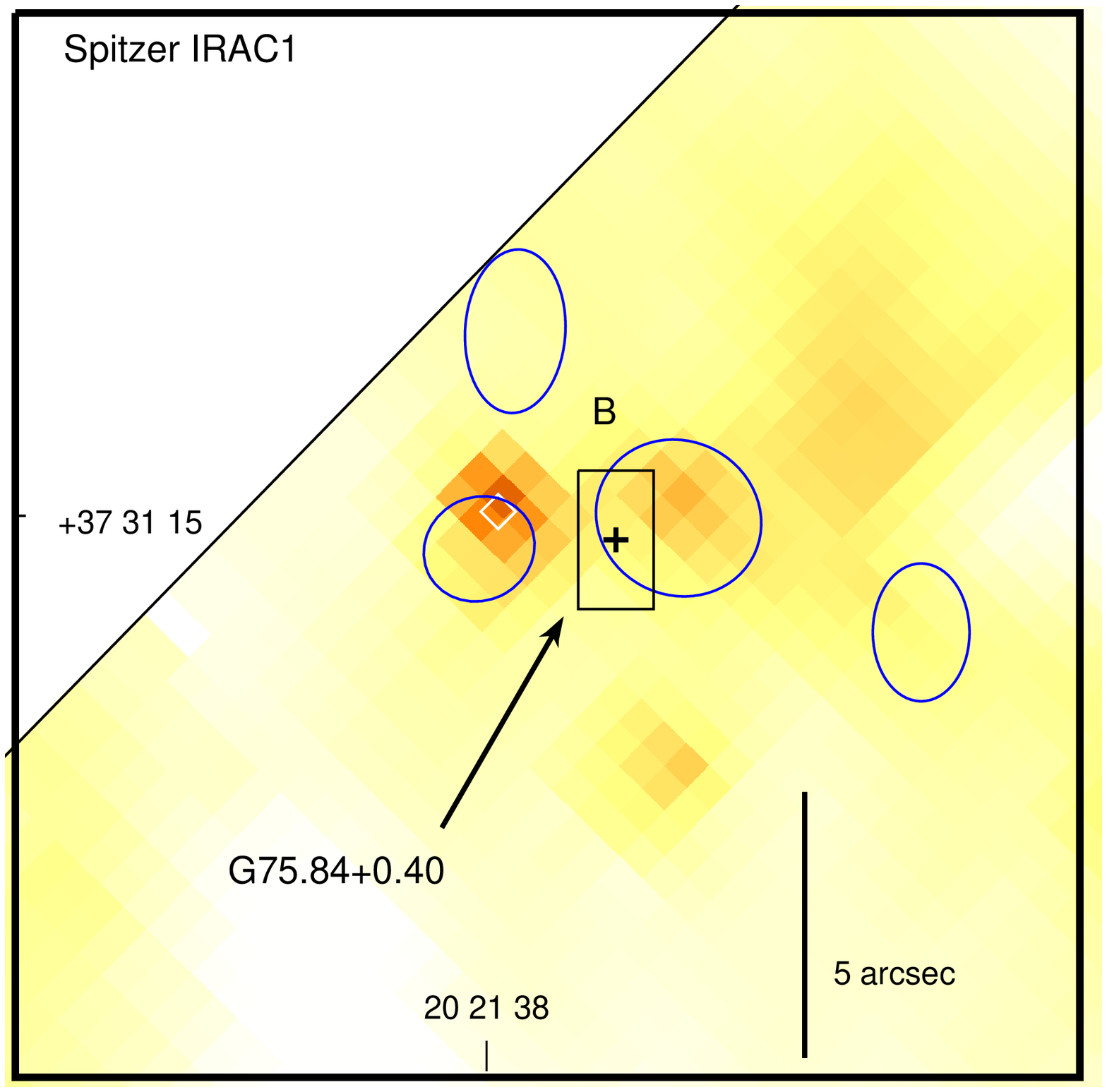}
\caption{{\em Top}:~{\em Spitzer} IRAC1 (3.6 $\mu$m) image of a 1$'$ $\times$ 1$'$
 region near G75.77$+$0.34 (AOR 18913536) displayed on log intensity scale.
{\em Chandra} source position error ellipses (3$\sigma$) are shown in 
black (ObsId 9914) and blue (ObsId 18083). The $\times$ symbols mark selected UGPS
near-IR source  positions, including those of {\em Chandra} counterparts (Table 2). 
Rectangles are position error boxes for Westerbork radio continuum sources 
G75.77$+$0.34A and B (M73).
The diamond overlapping peak A marks the  position of 2MASS J202142.13$+$372554.3
and the blue cross southwest of peak A is the VLA source position of Garay et al. (1993).
A faint IRAC source lies within
the radio position error box of G75.77$+$0.34A but not B. 
The position of UGPS J202140.54$+$372548.4 near peak B is marked ($\times$).
Coordinate tick marks are J2000.
~{\em Bottom}:~Same as above but for a 20$''$ $\times$ 20$''$ region near
G75.84$+$0.40B. A faint IRAC source overlaps the radio position error box
of peak B as does  a {\em Chandra} source (source 4 in Table 3). A second IRAC source with a 
2MASS and {\em Chandra} counterpart (source 6 in Table 3) lies about 2$''$.6 east of radio peak B.
Radio peak A fell outside the IRAC1 field-of-view.
}
\end{figure}

%% file: f5.tex
\begin{figure}
\figurenum{5}
\includegraphics*[width=8.0cm,angle=0]{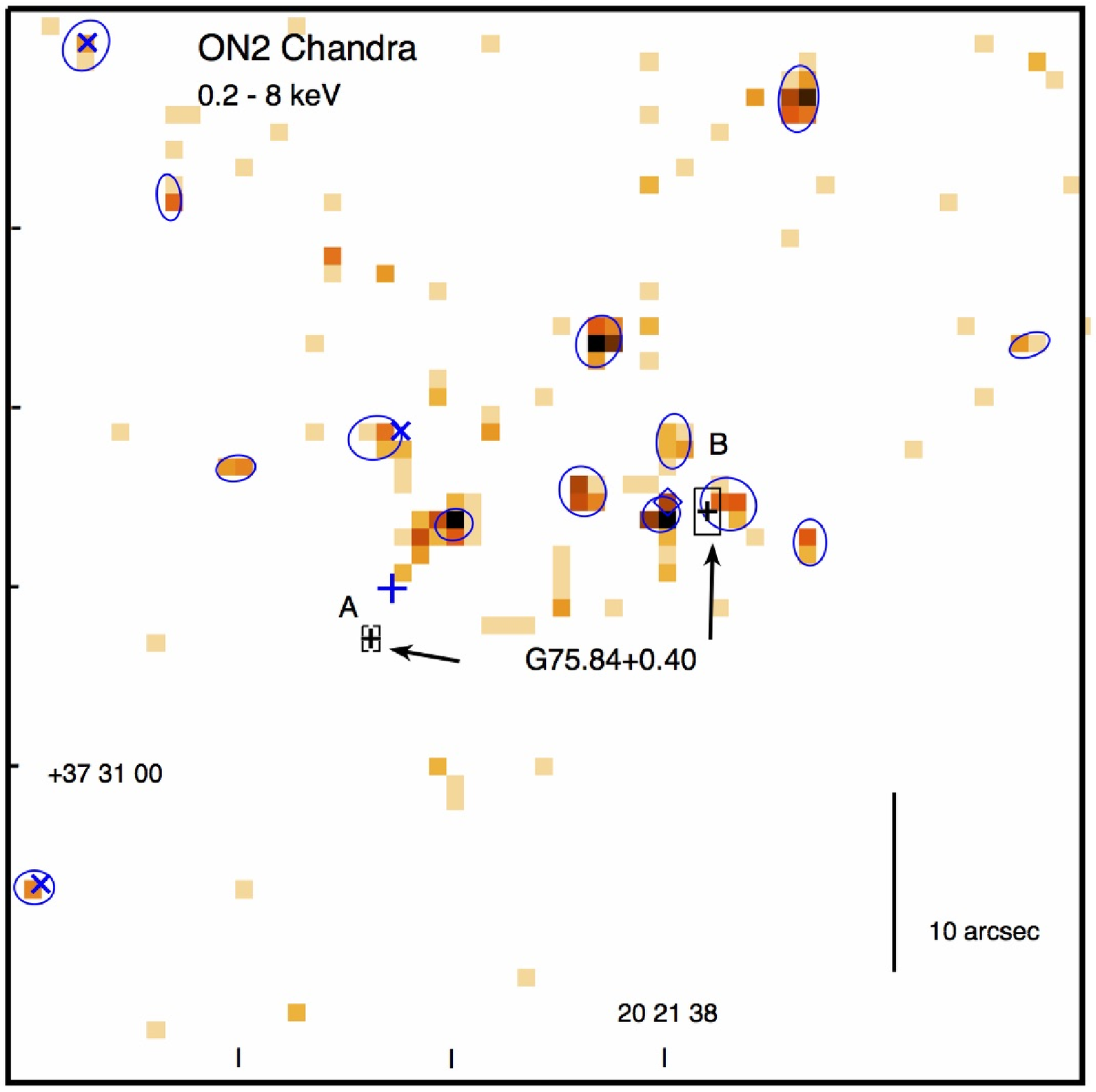} \\
\includegraphics*[height=7.5cm,angle=-90]{f5b.eps}
\caption{{\em Top}:~{\em Chandra} ACIS-I image (0.2 - 8 keV) of a
1$'$ $\times$ 1$'$ region near G75.84$+$0.40 (ObsId 18083). 
{\em Chandra} source position error ellipses (3$\sigma$) are shown.  
Near-IR counterparts (Table 3) are marked as a diamond (2MASS) and
$\times$ symbols (UGPS).  Rectangles are position error boxes for Westerbork
radio continuum sources G75.84$+$0.40A and B (M73).
The  position error ellipse of {\em Chandra} source CXO J202137.69$+$373115.1 
overlaps that of radio source G75.84$+$0.40B and a brighter {\em Chandra} source
CXO J202138.01$+$373114.5 with a 2MASS counterpart lies slightly further to the
east at an offset of 2$''$.6 from radio peak B.
There is no X-ray source at the position of Westerbork peak G75.84$+$0.40A
or at the VLA peak (shown as a blue $+$ symbol) of Garay et al. (1993).
Coordinate tick marks are J2000. Log intensity scale.
{\em Bottom}:~Binned {\em Chandra} ACIS-I spectra histograms of the two X-ray sources
lying near radio peak G75.84$+$0.40B. Spectral fits of both sources with isothermal 
models are summarized in Table 4.  Si XIII emission may be contributing to
the strong feature near 1.85 keV in CXO J202138.01$+$373114.5.
}
\end{figure}

%% file: f6.tex
\begin{figure}
\figurenum{6}
\includegraphics*[width=7.0cm,angle=0]{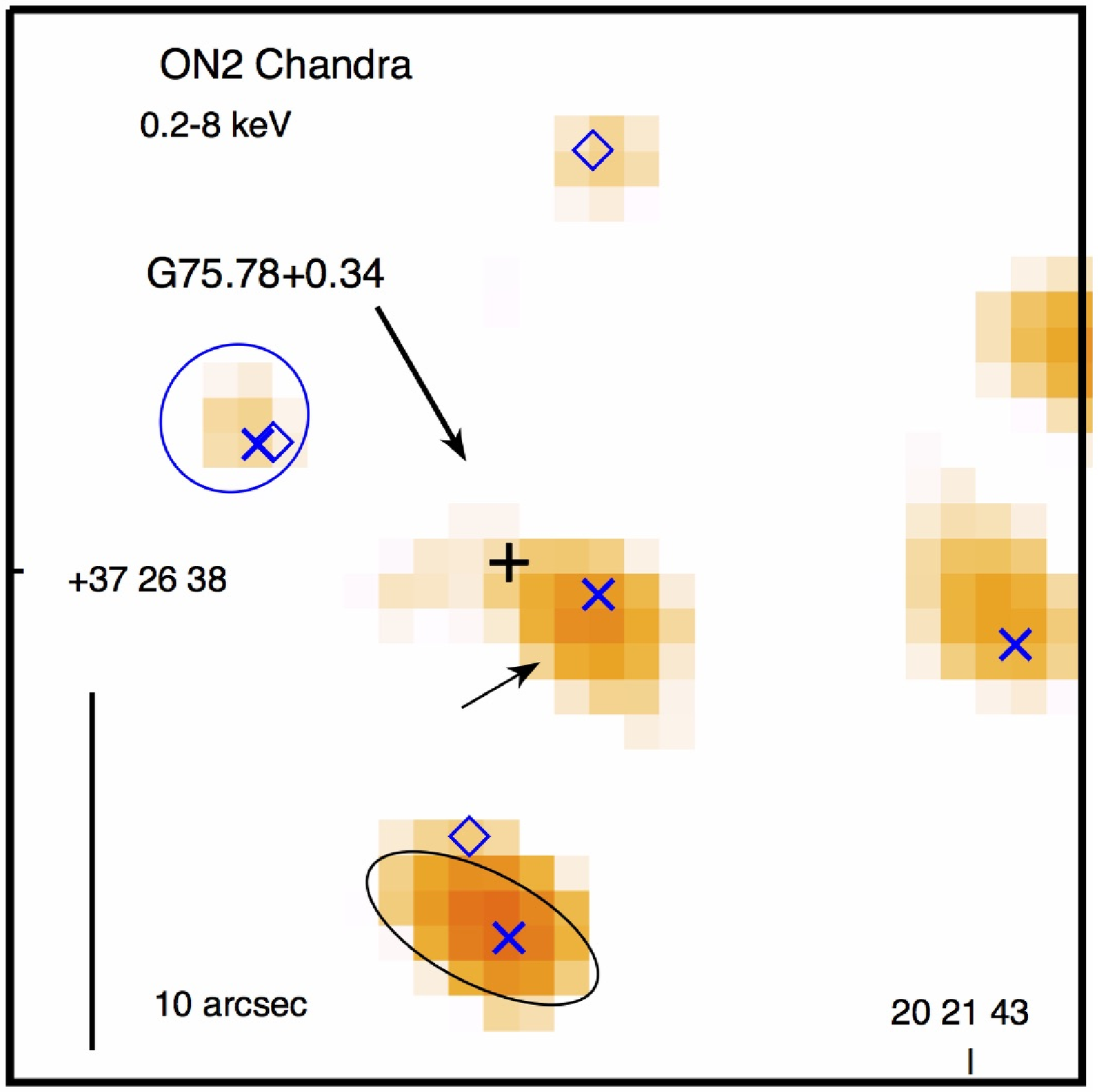}
\includegraphics*[width=7.0cm,angle=0]{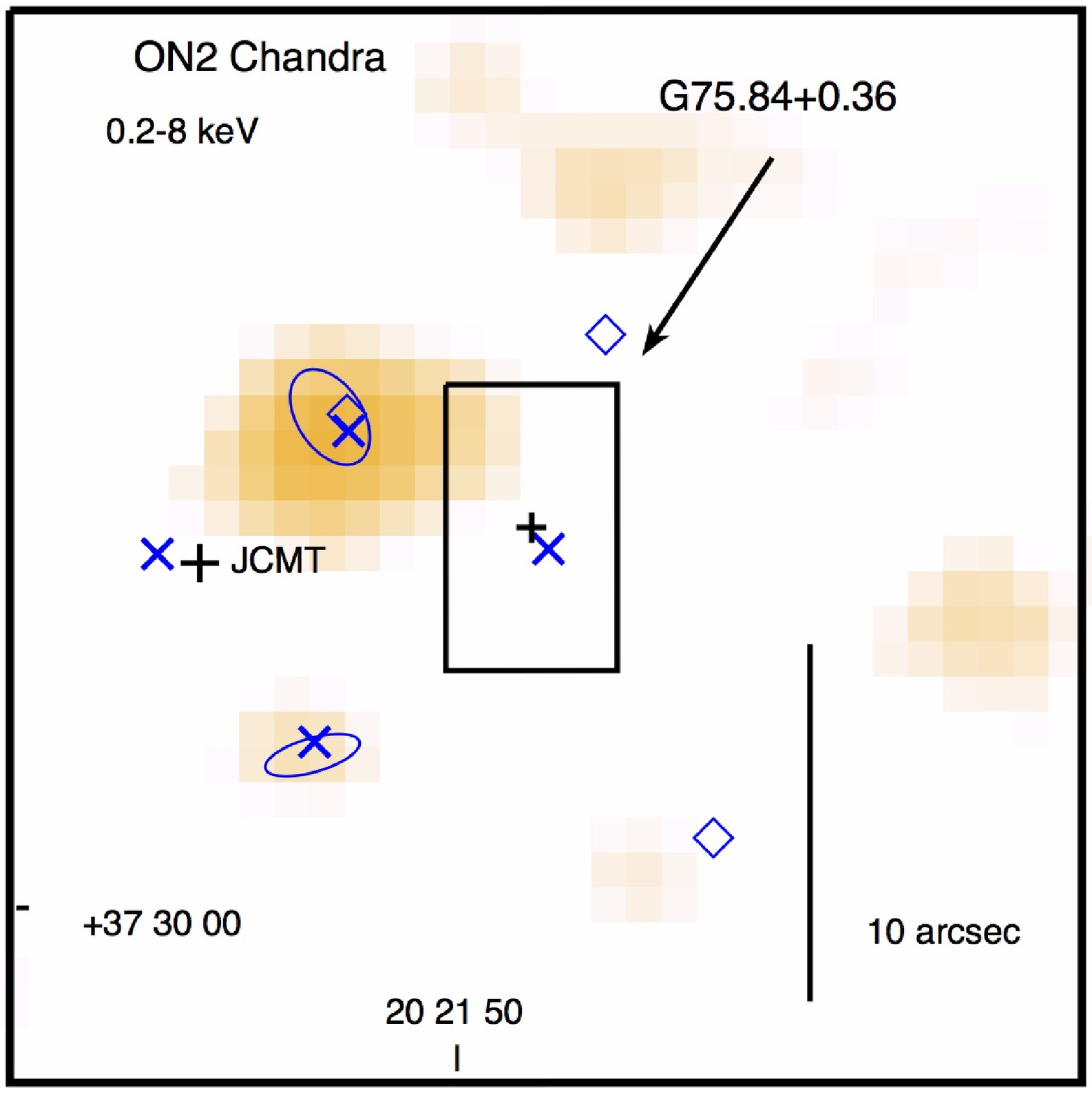}
\caption{Zoomed mosaiced {\em Chandra} ACIS-I images (0.2 - 8 keV)
of 30$''$ $\times$ 30$''$ regions near G75.78$+$0.34 (ucHII) and 
G75.84$+$0.36. The images have been lightly Gaussian-smoothed.
{\em Chandra} source position error ellipses (3$\sigma$) are shown 
in black for ObsId 9914 and blue for ObsId 18083.
Positions of selected near-IR UGPS ($\times$) and
2MASS (diamonds) sources are shown.
Log intensity scale. Coordinate tick marks are J2000. 
{\em Left}:~G75.78$+$0.34 region.
Cross at center marks the VLA radio
position J202144.08$+$372638.6 (Shepherd et al. 1997). 
The X-ray peak marked  with a short arrow (CXO J202143.92$+$372636.6) 
was not detected by the {\em wavdetect} source detection algorithm.  
{\em Right}:~G75.84$+$0.36 region. 
Cross ($+$) and rectangular box mark the radio position and 
its uncertainty from Matthews \& Spoelstra (1983). 
The JCMT continuum source position ($+$) is
from Di Francesco et al. (2008). There is no X-ray source
at the radio or JCMT positions. The nearest X-ray source
CXO J202150.30$+$373014.09 (= UGPS J202150.25$+$373013.7 =
2MASS J202150.25$+$373014.1) lies 6$''$.5 northeast of the radio
position. It was detected in both {\em Chandra} observations
and is variable with a count rate (0.2 - 8 keV) of 0.305 c ks$^{-1}$ 
in ObsId 9914 and 0.094 c ks$^{-1}$ in ObsId 18083.
}
\end{figure}

%% file: f7.tex
\begin{figure}
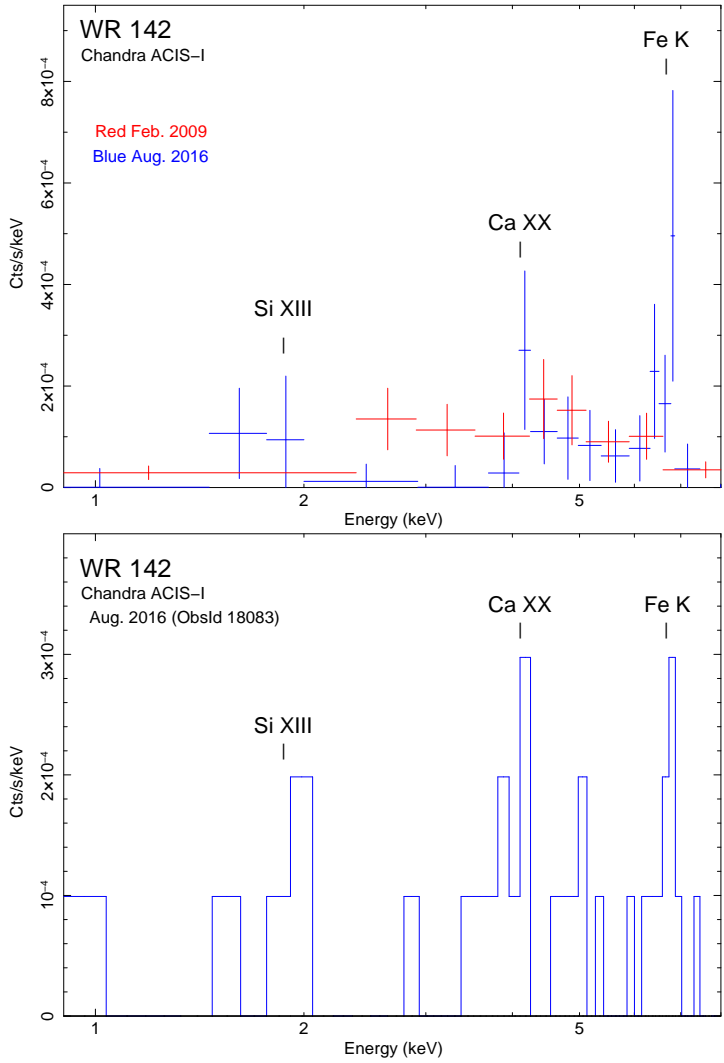

\figurenum{7}
\includegraphics*[height=10.0cm,width=7.0cm,angle=-90]{f7t.eps} \\
\includegraphics*[height=10.0cm,width=7.0cm,angle=-90]{f7b.eps}
\caption{{\em Top}:~Overlay of {\em Chandra} ACIS-I background-subtracted spectra of WR 142  rebinned to
           a minimum of 5 counts per bin (ObsId 9914; Feb. 2009; 46 net counts)
           and 3 counts per bin (ObsId 18083; Aug. 2016; 26 net counts). 
           Weak line emisson may be present from Si XIII (1.855 - 1.865 keV),
           Ca XX (4.107 keV) and the Fe K complex (Fe XXV; 6.67 - 6.68 keV)
           in the Aug. 2016 observation.
           {\em Bottom}:~A histogram plot of the WR 142 ACIS-I spectrum (ObsId 18083)
           rebinned to a minimum of 3 counts per bin showing possible faint emission lines.
           Error bars omitted for clarity.
}
\end{figure}

%% file: f8.tex
\begin{figure}
\figurenum{8}
\includegraphics*[width=7.0cm,angle=0]{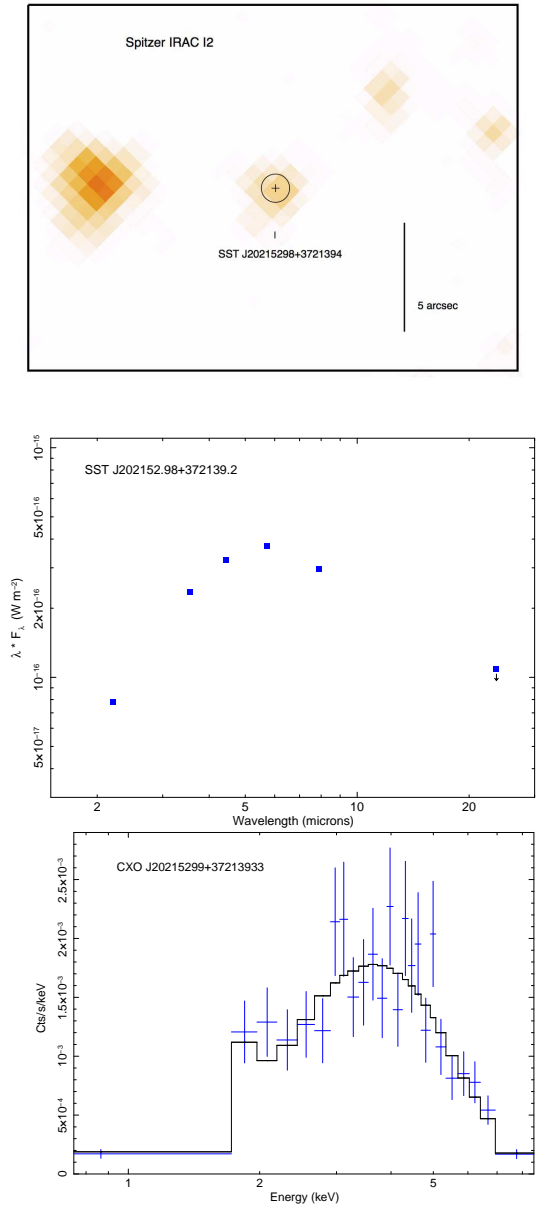} \\
\includegraphics*[height=7.0cm,angle=-90]{f8m.eps} \\
\includegraphics*[height=7.0cm,angle=-90]{f8b.eps}
\caption{{\em Top}:~{\em Spitzer} IRAC 4.5 $\mu$m image (AOR 18913536) of the region 
near the brightest X-ray source detected in ON 2, CXO J202152.99$+$372139.33. Cross marks the 
{\em Chandra} X-ray position centroid and the X-ray position 3$\sigma$ 
error circle has a radius of 1$''$.3. N is up, E is left.
{\em Middle}:~IR spectral energy distribution of {\em Spitzer} source
 SSTSL2 J202152.98$+$372139.2 (top panel) based on data from 
 UGPS (K = 17.63), IRAC, and the MIPS 24 $\mu$m 3$\sigma$ upper limit.
{\em Bottom}:~{\em Chandra} ACIS-I spectrum of the bright heavily-absorbed
source CXO J202152.99$+$372139.33 (ObsId 9914) rebinned to a minimum of 20 counts per bin.
The solid line is a best-fit absorbed power-law model with a photon power-law
index $\Gamma_{ph}$ = 1.15 (Sec. 3.7). 
}
\end{figure}

%% file: f9.tex
\begin{figure}
\figurenum{9}
\includegraphics*[width=7.0cm,angle=0]{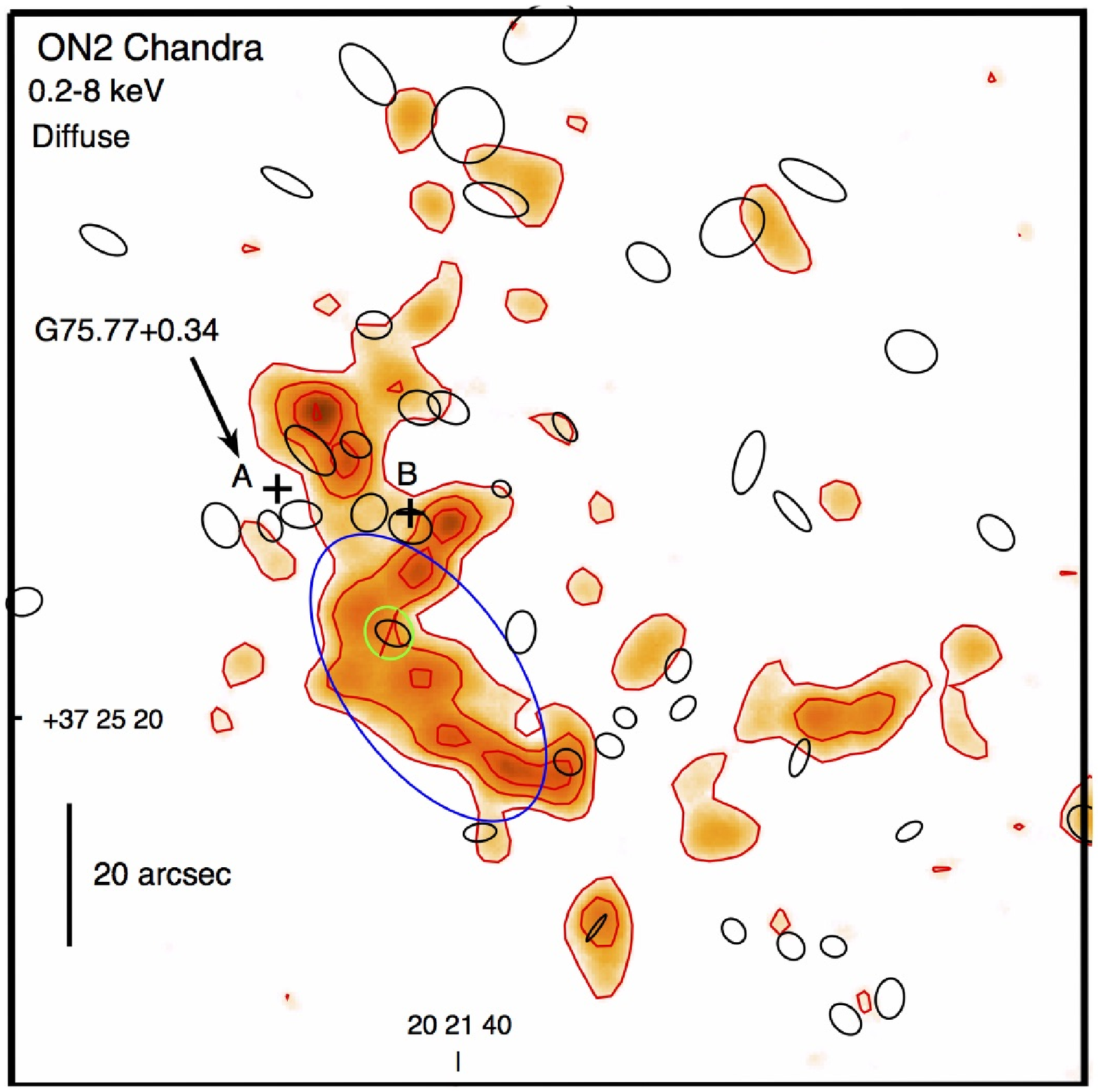}
\includegraphics*[width=7.0cm,angle=0]{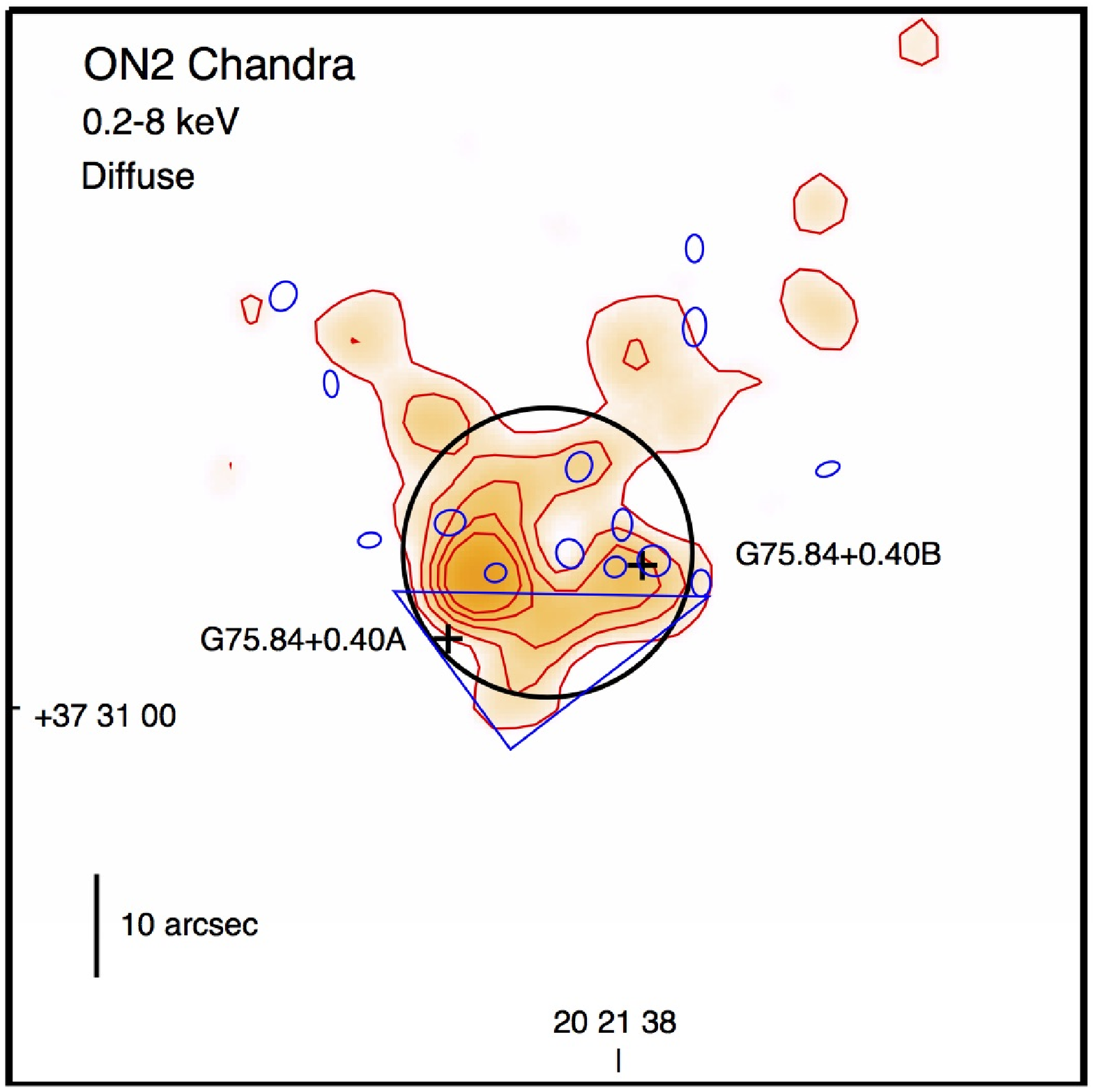}
\includegraphics*[height=7.0cm,angle=-90]{f9bl.eps}
\hspace{3.0cm}
\includegraphics*[height=7.0cm,angle=-90]{f9br.eps}
\caption{{\em Top}:~Smoothed {\em Chandra} ACIS-I 0.2-8 keV images 
near the HII region G75.77$+$0.34 (left; ObsId 9914)
and G75.84$+$0.40 (right; ObsId 18083) showing  faint diffuse emission.
The images were generated by deleting emission inside X-ray point sources
(small ellipses) and filling in the expunged regions 
with adjacent background emission. The region inside the large blue 
ellipse (semi-axes 23$''$ $\times$ 12$''$) in the G75.77$+$0.34 image 
was used to extract the diffuse spectrum, excluding the small region 
inside the green circle where a point source was detected.  
For G75.84$+$0.40, diffuse spectra were extracted
inside the large r=14$''$ circle  and the smaller blue triangle
for comparison.  The images have been Gaussian-smoothed using an 11-pixel (5$''$.4) kernel
to bring out faint emission. Contours are at levels (2,3,4.5.6)$\sigma$,
where $\sigma$ is the background level in the smoothed image.
Crosses mark the radio continuum peak  positions of 
G75.77$+$0.34 A/B and G75.84$+$0.40 A/B (M73).
Coordinate tick marks are J2000.
{\em Botttom}:~Net diffuse spectra (black) binned to a minimum of
5 cts per bin and binned background spectra (red) extracted in source-free
regions on the same CCD. The nominal background has been subtracted from
the diffuse spectra.
Left: G75.77$+$0.34 diffuse spectrum (160 net cts).~
Right: G75.84$+$0.40 diffuse spectrum (r=14$''$ circle, 131 net counts).
}
\end{figure}

%% file: f10.tex
\begin{figure}
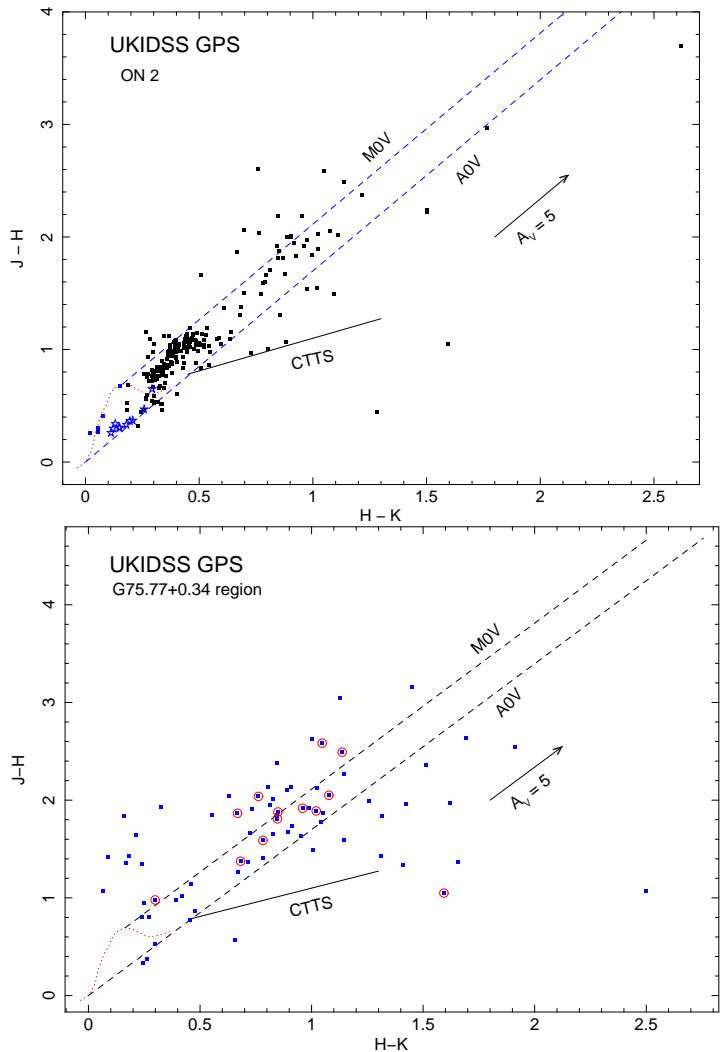

\figurenum{10}
\includegraphics*[width=7.0cm,angle=-90]{f10t.eps} \\
\includegraphics*[width=7.0cm,angle=-90]{f10b.eps}
\caption{
{\em Top}:~
UKIDSS GPS (UGPS) color-color diagram for near-IR counterparts within 1$''$ 
of  {\em Chandra} sources detected in ON 2. Only those UGPS sources with
detections in all three bands are shown. 
The blue star symbols show 2MASS colors for
Berkeley 87 members 87-3,4,13,24,25,26,29,32,65,94 listed in TF82.
Blue squares show 2MASS colors for stars listed in TF82 that
are determined to be foreground objects based on GAIA DR2 distances $<$ 900 pc:
~Berk 87-6,11,19,28,104.
The long dashed lines show the locus of normally-reddened 
M0V and A0V stars based on unreddened colors in 
Bessell \& Brett (1988) and the reddening law of Meyer et al. (1997).
The solid line is the locus of {\em unreddened} cTTS from Meyer et al. (1997).
Objects lying above the cTTS line and to the right of the A0V line
(Table 7) have colors consistent with reddened cTTS. 
{\em Bottom}:~UGPS color-color diagram for {\em all} UGPS sources with
detections in all three bands lying within a 1$'$ $\times$ 1$'$ region
surrounding the G75.77$+$0.34AB radio position (Fig. 3-top).
Sources circled in red were detected by {\em Chandra} 
(Table 2, excluding sources 7 and 15 which lack complete UGPS photometry).
The circled  source  at (1.59,1.05) is
UGPS J202141.01$+$372550.0, the counterpart of
CXO J202141.02$+$372550.2. Photometry of the object below the cTTS locus at
(0.656,0.571) may be contaminaed by a second source at an offset of 0$''$.4
}
\end{figure}

%% file: f11.tex
\begin{figure}
\figurenum{11}
\includegraphics*[width=12.0cm,angle=0]{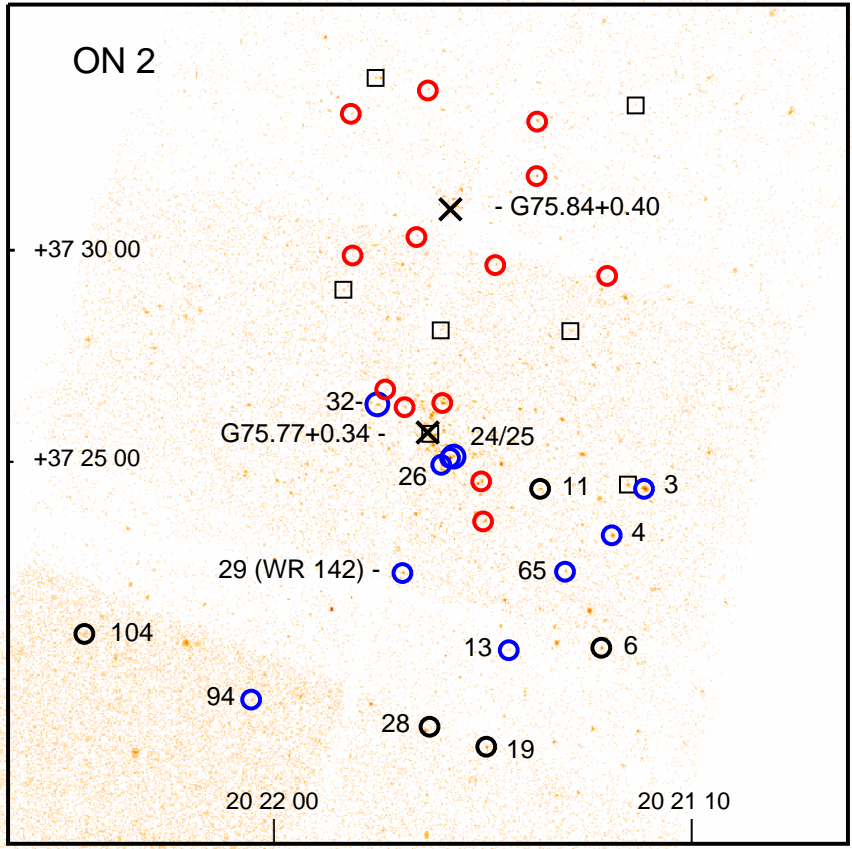}
\caption{{\em Chandra} ACIS-I ON 2 image (0.2 - 8 keV)
overlaid with positions of X-ray detected UGPS counterparts
lying to the right (redward) of the normally-reddened A0V line in 
Figure 10-top (see also Table 7). 
Red circles are cTTS candidates which lie on or above  the
unreddened cTTS locus.
Black squares lie below the cTTS locus
and are only marginally reddened and not strong cTTS candidates.
Numbered blue circles mark positions of some Berkeley 87 cluster 
members from the catalog of TF82 that were detected in 
X-rays (Sokal et al. 2010). Numbered black circles correspond 
to X-ray sources that were listed as candidate Berkeley 87 stars
by TF82 but are now determined to be foreground stars with
GAIA DR2 distances $<$900 pc (Berk 87-6,11,19,28,104).
The mean radio peak positions 
of G75.77$+$0.34 and G75.84$+$0.40 are marked ($\times$).
The bounding box has dimensions 20$'$ $\times$ 20$'$.
Coordinate tick marks are J2000.
}
\end{figure}